\documentclass[preprint,showpacs,preprintnumbers,amsmath,amssymb]{revtex4}
\usepackage{graphicx}
\usepackage{dcolumn}
\usepackage{bm}
\begin{document}
\title{Variation of the magnetic ordering in GdT$_2$Zn$_{20}$ (T= Fe, Ru, Os, Co, Rh and Ir) and its correlation with the electronic structure of isostructural YT$_2$Zn$_{20}$}
\author{Shuang Jia$^{1,2}$, Ni Ni$^{1,2}$, G. D. Samolyuk$^{1}$, A. Safa-Sefat$^{1}$, K. Dennis$^{1}$, Hyunjin Ko$^{1,3}$, G. J. Miller$^{1,3}$, S. L. Bud'ko$^{1,2}$, P. C. Canfield$^{1,2}$}
\affiliation{$^{1}$Ames Laboratory, USDOE\\ $^{2}$Department of Physics and Astronomy\\ $^{3}$Department of Chemistry,\\ Iowa State University\\Ames, Iowa 50011, USA\\}

\begin{abstract}
Magnetization, resistivity and specific heat measurements were performed on the solution-grown, single crystals of six GdT$_2$Zn$_{20}$ (T = Fe, Ru, Os, Co, Rh and Ir) compounds, as well as their Y analogues.
For the Gd compounds, the Fe column members manifest a ferromagnetic (FM) ground state (with an enhanced Curie temperature, $T_{\mathrm{C}}$, for T = Fe and Ru), whereas the Co column members manifest an antiferromagnetic (AFM) ground state.
Thermodynamic measurements on the YT$_2$Zn$_{20}$ revealed that the enhanced $T_{\mathrm{C}}$ for GdFe$_2$Zn$_{20}$ and GdRu$_2$Zn$_{20}$ can be understood within the framework of Heisenberg moments embedded in a nearly ferromagnetic Fermi liquid.
Furthermore, electronic structure calculations indicate that this significant enhancement is due to large, close to the Stoner FM criterion, transition metal partial density of states at Fermi level, whereas the change of FM to AFM ordering is associated with filling of electronic states with two additional electrons per formula unit.
The degree of this sensitivity is addressed by the studies of the pseudo-ternary compounds Gd(Fe$_x$Co$_{1-x}$)$_2$Zn$_{20}$ and Y(Fe$_x$Co$_{1-x}$)$_2$Zn$_{20}$ which clearly reveal the effect of $3d$ band filling on their magnetic properties. 

\end{abstract}


\maketitle
\section{Introduction}

Magnetism of rare earth intermetallics, determined by the interaction between $4f$ local moments and conduction electrons, especially the $d$-band conduction electrons of transition metals, has been of interest to physicists for several decades.\cite{franse_magnetic_1993, szytula_handbook_1994}
Recently, studies of the dilute, rare earth bearing, intermetallic compounds, RT$_2$Zn$_{20}$ (R = rare earth, T = transition metal in Fe, Co or neighboring groups), revealed varied, exotic magnetic properties.\cite{jia_nearly_2007, torikachvili_six_2007, jia_GdY_2007}
Containing less than 5 at. \% rare earth ions which, although dilute, fully occupy a unique crystallographic site, these compounds allow for the study of local and hybridizing moment magnetism in a regime that approaches the single ion limit while preserving periodicity.
Previous studies of these compounds have indicated that they can serve as model systems for a variety of physical phenomenon ranging from a nearly ferromagnetic Fermi liquid (NFFL) (YFe$_2$Zn$_{20}$ and LuFe$_2$Zn$_{20}$),\cite{jia_nearly_2007} to greatly enhanced ferromagnetic (FM) order in GdFe$_2$Zn$_{20}$,\cite{jia_nearly_2007, jia_GdY_2007} all the way to heavy fermion ground states in YbT$_2$Zn$_{20}$ (T = Fe, Ru, Os, Co, Rh and Ir).\cite{torikachvili_six_2007}

\begin{figure}
  \begin{center}
  \includegraphics[clip, width=0.45\textwidth]{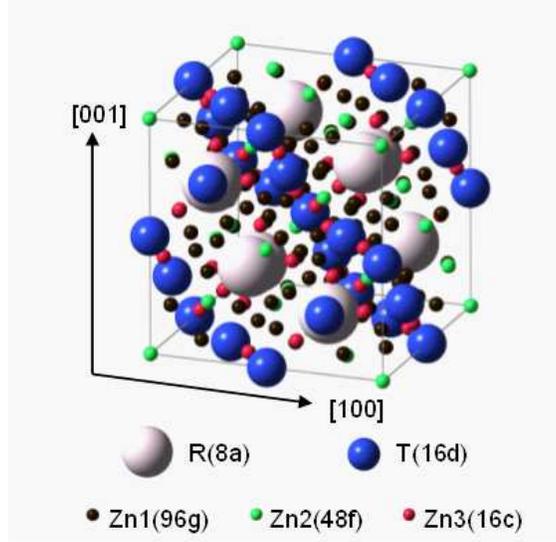}\\
  \caption{(color online) The crystal structure of RT$_2$Zn$_{20}$. The thin lines outline the cubic unit cell. The unit cell dimension, $a$, is approximately 14 {\AA} for the RT$_2$Zn$_{20}$ families.}
  \label{Fig1UC}
  \end{center}
\end{figure}

The RT$_2$Zn$_{20}$ series of compounds were discovered in 1997 by Nasch et al.\cite{nasch_ternary_1997}
These compounds assume the isostructural, cubic, CeCr$_2$Al$_{20}$ structure\cite{kripyakevich_RCr2Al20_1968, thiede_euta2al20_1998, moze_crystal_1998}, in which the R and T ions each occupy their own single, unique crystallographic site with cubic and trigonal point symmetry respectively, and the Zn ions have three unique crystallographic sites (Fig. \ref{Fig1UC}).
The coordination polyhedra for R and T are fully comprised of Zn, meaning that there are no R-R, T-T or R-T nearest neighbors and the shortest R-R spacing is $\sim$~6 {\AA}.
The nearest-neighbor and next-nearest-neighbor shells of the R are all Zn, forming an all Zn Frank-Kasper polyhedron around, and isolating the site.\cite{nasch_ternary_1997}
RT$_2$Zn$_{20}$ compounds had been found to form for T = Fe, Ru, Co and Rh, but no thermodynamic or transport property measurements were reported.
As part of this study we have extended the range of known RT$_2$Zn$_{20}$ compounds to T = Os and Ir series.

In rare earth containing series of intermetallic compounds, R = Gd members give the clearest indication of the strength and sign of the magnetic interaction, without any complications associated with crystalline electric field splitting of the Hund's rule ground state multiplet.
In order to better understand the RT$_2$Zn$_{20}$ series of compounds, in this paper we examine the thermodynamic and transport properties of six GdT$_2$Zn$_{20}$ (T = Fe, Ru, Os, Co, Rh and Ir) compounds as well as their R = Y analogues.
We found FM transitions in the iron column members (with enhanced $T_{\mathrm{C}}$ values for T = Fe and Ru) and low temperature antiferromagnetic (AFM) transitions in the cobalt column members.
Consistent with these results, we also found enhanced paramagnetism in the T = Fe and Ru of YT$_2$Zn$_{20}$ analogues.
For GdFe$_2$Zn$_{20}$ and GdRu$_2$Zn$_{20}$, magnetization measurements under hydrostatic pressure indicated that their enhanced FM transitions are not primarily associated with a steric effect.
A model of Heisenberg moments embedded in a NFFL can be proposed as a way to understand the enhanced FM transitions.
Band structure calculations were employed to explain that the remarkable differences in magnetic ordering for different transition metal members are a result of different $d$-band filling.
In order to test this further, a series of pseudo-ternary compounds Y(Fe$_x$Co$_{1-x}$)$_2$Zn$_{20}$ and Gd(Fe$_x$Co$_{1-x}$)$_2$Zn$_{20}$ were made, characterized and found to manifest a clear, systematic and comprehensible evolution from normal, to nearly FM, metal, and from AFM state to high temperature FM state, respectively, associated with a change of the $d$-band filling as $x$ varies from 0 to 1.   
 
\section{Experimental Methods and Calculation Details}

Single crystals of RT$_2$Zn$_{20}$ (R = Gd, Y; T = Fe, Co, Ru, Rh, Os and Ir) were grown from a Zn-rich self flux.\cite{canfield_growth_1992, jia_nearly_2007}
The initial concentration of starting elements (R:T:Zn) were 2: 4: 96 (T = Fe and Co), 1: 2: 97 (T = Ru, Rh), 1: 0.5: 98.5 (T = Os), and 0.75: 1.5: 97.75 (T = Ir).
High purity, constituent elements were placed in alumina crucibles and sealed in quartz tubes under approximately 1/3 atmosphere of high purity Ar. Then the ampules were heated up to 1000 $^\circ$C (T = Fe and Co), 1150 $^\circ$C (T = Ru), 1100 $^\circ$C (T = Rh), 1150 $^\circ$C (T = Os and Ir), and cooled down to 600 $^\circ$C, 850 $^\circ$C, 700 $^\circ$C, 750 $^\circ$C respectively, at which point the remaining liquid was decanted.
The cooling rates were 5 $^\circ$C/hr (T = Fe, Co, Ru, Rh), 4 $^\circ$C/hr (T = Os), and 2.5 $^\circ$C/hr (T = Ir).
Growths such as these often had only 2--3 nucleation sites per crucible and yielded crystals with typical dimensions of $7\times 7\times 7$ $mm^3$ or larger except for the Os compounds, which were significantly smaller (1--2 $mm$ on one side).
Residual flux and/or oxide slag on the crystal surfaces was removed by using diluted acid (0.5 vol. \% HCl in H$_2$O for T = Fe, Co or 1 vol. \% acetic acid in H$_2$O for T = Ru, Rh, Os and Ir).
The samples were characterized by room temperature, powder X-ray diffraction measurements using Cu $K_\alpha $ radiation with Si ($a=5.43088$ \AA) as an internal standard. 
The lattice constants were obtained by using the Rietica, Rietveld refinement program.

Subsequent single crystal X-ray analyses were made using a STOE image plate diffractometer with Mo $K_\alpha $ radiation using the supplied STOE software\cite{stoe_area-software_2002}.
The data were adjusted for Lorentz and polarization effects, and a numerical absorption correction was done.
The structural solutions were refined by full-matrix least-squares refinement using Bruker SHELXTL 6.1 software package\cite{sheldrick_2000}.
The atomic disorder in the crystals was checked by refining site occupancies. 

The magnetization measurements under hydrostatic pressure were preformed in a piston-cylinder clamp-type pressure cell, made out of non-magnetic Ni-Co alloy MP35N, in the Quantum Design superconducting quantum interface device (SQUID) magnetometers.
Pressure was generated in a Teflon capsule filled with 50:50 mixture of n-pentane and mineral oil.
The pressure dependent, superconducting transition temperature of 6-N purity Pb was employed to determine the pressure at low temperatures.\cite{Pb_pressure}
The pressure cell design allows for the routine establishment of pressures in excess of 8 kbar at low temperatures.\cite{budko_pressure}

Measurements of the electrical resistivity were made by using a standard AC, four-probe technique.
The samples were cut as bars, which typically had length 2--3 $mm$, parallel to the crystallographic  [110] direction. AC electrical resistivity measurements were taken on these bars with $f=16$ Hz, $I = 0.5$--$0.3$ mA in Quantum Design physical properties measurement system, PPMS-14 and PPMS-9 instrument ($T = 1.85$--$310$ K).
Temperature dependent specific heat measurements were also performed by using the heat capacity option of these Quantum Design instruments.
DC magnetization was measured in Quantum Design SQUID magnetometers, in applied field $\leq 55$ kOe or $70$ kOe and in the temperature range from 1.85 K to 375 K.

In general, when making magnetization measurements on FM samples, some attention must be paid to the effects of demagnetizing fields.\cite{chikazumi_physics_1997}
However, this correction is small in the case of GdT$_2$Zn$_{20}$ because of the diluted nature of the magnetic moments.
Considering that the magnetization is mainly from the eight Gd$^{3+}$ ions per unit cell, one estimates the maximum demagnetizing field as:

\begin{equation}
D_{m}(Max)=4\pi \frac{8\times 7~\mu _B}{(14~\AA)^{3}}=2380~Oe.
\label{eqn:1}
\end{equation}

Experimentally, in the measurements of magnetization isotherms near $T_{\mathrm{C}}$, the demagnetizing field can introduce an error of $T_{\mathrm{C}}$ for plate-like shaped samples.
To avoid this error, rod-like samples were measured with the applied magnetic field along their long axis.
This minimized the demagnetizing factor and thereby the demagnetizing field.

The electronic structure was calculated using the atomic sphere approximation, tight binding, linear muffin-tin orbital (TB-LMTO-ASA) method\cite{andersen_linear_1975, andersen_explicit_1984} with the experimental values of the lattice parameters and atomic positions from this work. 
The exchange-correlation term was calculated both within the local-spin-density approximation (LSDA) which was parameterized according to von Barth-Hedin\cite{Barth_local_1972}, and the generalized gradient approximation (GGA) with the Perdew-Burke-Ernzerhof functional\cite{perdew_generalized_1996}.
A mesh of 16 $\vec{k}$ points in the irreducible part of the Brillouin zone (BZ) was used. 
The $4f$ electrons of the Gd atoms were treated as polarized core states.
Despite its apparent simplicity, this approach reproduces the electronic and magnetic properties of rare earth in good agreement with experiment.\cite{perlov_rare, turek_ab}
In order to reproduce the AFM ordering in GdCo$_2$Zn$_{20}$ the magnetic moments of two Gd atoms in the unit cell were aligned in opposite direction.
\section{Results and Analysis}

\subsection{\label{sec:A}Structure refinements}

\begin{figure}
  \begin{center}
  \includegraphics[clip, width=0.45\textwidth]{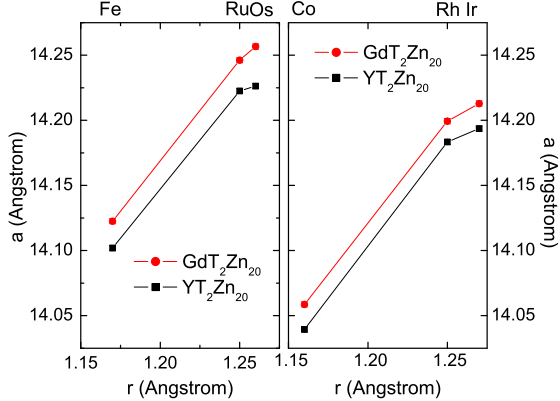}\\
  \caption{(color online) The lattice constants ($a$) of GdT$_2$Zn$_{20}$ and YT$_2$Zn$_{20}$ versus the radius of the free, trivalent transition metal ion ($r$)\cite{periodic_table}.}
  \label{Fig2lattice}
  \end{center}
\end{figure}

Shown in Fig. \ref{Fig2lattice}, the lattice parameters, determined by the refinement of powder X-ray diffraction, increase as the transition metal varies from $3d$ to $5d$ for both of GdT$_2$Zn$_{20}$ and YT$_2$Zn$_{20}$.
The error bars, smaller than the symbols in the plot, were estimated from the standard variation of multiple measurement results on one batch of sample.
In addition to the refinement of powder X-ray diffraction, the crystallographic atomic site occupancies and positions were refined using single crystal X-ray data on the crystals of GdFe$_2$Zn$_{20}$ and GdRu$_2$Zn$_{20}$.
Shown in Table \ref{table1}, both compounds were found to be fully or very close to fully stoichiometric.
The atomic site positions are very close to the isostructural compounds reported before\cite{nasch_ternary_1997}.
It should be noted, though, that the similar atomic number values for Zn and Fe made it difficult to resolve possible mixed site occupancies.

\begin{table}
\caption{\label{table1}	Atomic coordinates and refined site occupancies for GdFe$_2$Zn$_{20}$ and GdRu$_2$Zn$_{20}$; each of the unique crystallographic sites were refined individually. }
\begin{ruledtabular}
\begin{tabular}{cccccc}
Atom & Site & Occupancy & $x$ & $y$ & $z$\\
\hline
\multicolumn{6}{c}{GdFe$_2$Zn$_{20}$}\\
Gd & $8a$ & 1.013(12) & 0.125 & 0.125 & 0.125\\
Fe & $16b$ & 1.01(2) & 0.5 & 0.5 & 0.5\\
Zn1 & $96g$ & 0.993(7) & 0.0587(1) & 0.0587(1) & 0.3266(1)\\
Zn2 & $48f$ & 0.997(9) & 0.4893(1) & 0.1250 & 0.1250\\
Zn3 & $16c$ & 1.006(18) & 0 & 0 & 0\\
\hline
\multicolumn{6}{c}{GdRu$_2$Zn$_{20}$}\\
Gd & $8a$ & 1.026(9) & 0.125 & 0.125 & 0.125\\
Ru & $16b$ & 1.030(11) & 0.5 & 0.5 & 0.5\\
Zn1 & $96g$ & 0.988(5) & 0.0589(1) & 0.0589(1) & 0.3260(1)\\
Zn2 & $48f$ & 1.000(8) & 0.4888(1) & 0.1250 & 0.1250\\
Zn3 & $16c$ & 0.962(15) & 0 & 0 & 0\\
\end{tabular}
\end{ruledtabular}
\end{table}

\subsection{\label{sec:B}GdT$_2$Zn$_{20}$(T = Fe, Co, Ru, Rh, Os and Ir)}

\begin{figure}
  \begin{center}
  \includegraphics[clip, width=0.45\textwidth]{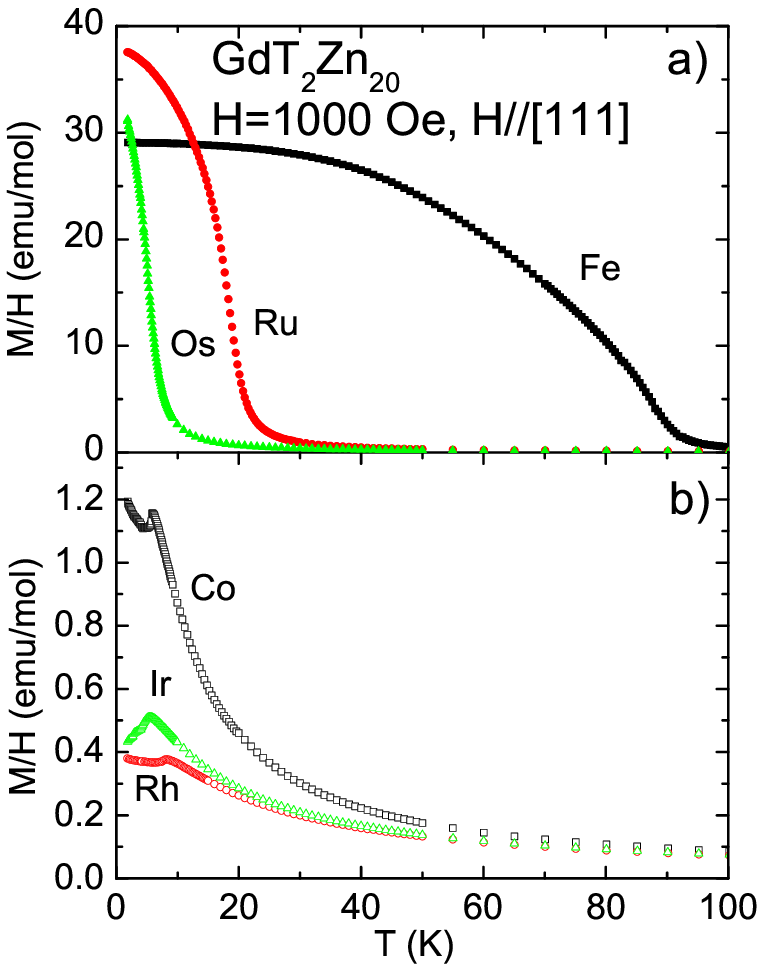}\\
  \caption{(color online) Temperature dependent magnetization of GdT$_2$Zn$_{20}$, divided by applied field $H = 1000$~Oe.}
  \label{Fig3MTall}
  \end{center}
\end{figure}

Before discussing each of the GdT$_2$Zn$_{20}$ compounds separately, an overview of their temperature and field dependent magnetization serves as a useful point of orientation.
In Fig. \ref{Fig3MTall} the temperature dependent magnetization ($M$) divided by applied field ($H$) reveals the primary difference between the Fe column members of this family and the Co column members.
For T = Fe, Ru and Os there is an apparent FM ordering (with remarkably high and moderately high values of $T_{\mathrm{C}}$ for T = Fe and Ru respectively), whereas for T = Co, Rh and Ir there is an apparent, low temperature AFM ordering. 

\begin{figure}
  \begin{center}
  \includegraphics[clip, width=0.45\textwidth]{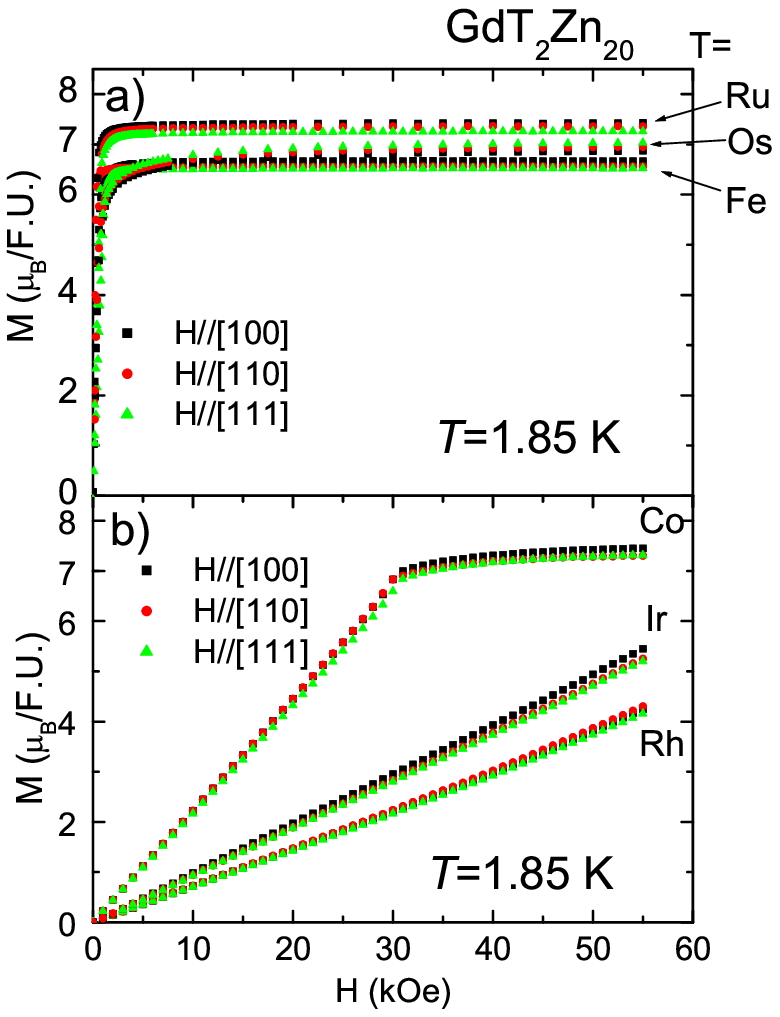}\\
  \caption{(color online) Field dependent magnetization of GdT$_2$Zn$_{20}$ at 1.85~K.}
  \label{Fig4MHall}
  \end{center}
\end{figure}

The nature of the ordering is further confirmed by the low temperature, magnetization isotherms presented in Fig. \ref{Fig4MHall}.
It should be noted that for each of the six GdT$_2$Zn$_{20}$ compounds, the 1.85 K magnetization isotherms, measured with the applied field along [100], [110], [111] crystallographic directions, were found to be isotropic to within less than 5 \%.
This magnetic isotropy is not unexpected in the Gd-based intermetallics, in which the magnetism is mainly due to the pure spin contribution of the $4f$ shell of Gd$^{3+}$.
For T = Fe, Ru and Os the magnetization is representative of a FM-ordered state with a rapid rise and saturation of the ordered moment in a field of the order of the estimated demagnetizing field (magnetic domain wall pinning being low in these single crystalline samples).
For T = Co, Rh and Ir the field dependent magnetization data are consistent with AFM-ordered states that can be field stabilized to fully saturated states in large enough applied fields.
This fully saturated state is observed for GdCo$_2$Zn$_{20}$ associated with a spin-flop transition near $H=31$ kOe, whereas the maximum magnetic field in the equipment used (55 kOe) could not saturate the magnetic moment of the GdRh$_2$Zn$_{20}$ and GdIr$_2$Zn$_{20}$ samples. 
The measured saturated moments for T = Fe, Ru, Os and Co samples are clustered around the Hund's rule ground state value of Gd$^{3+}$, $7~\mu_B$. 

\begin{figure}
  \begin{center}
  \includegraphics[clip, width=0.45\textwidth]{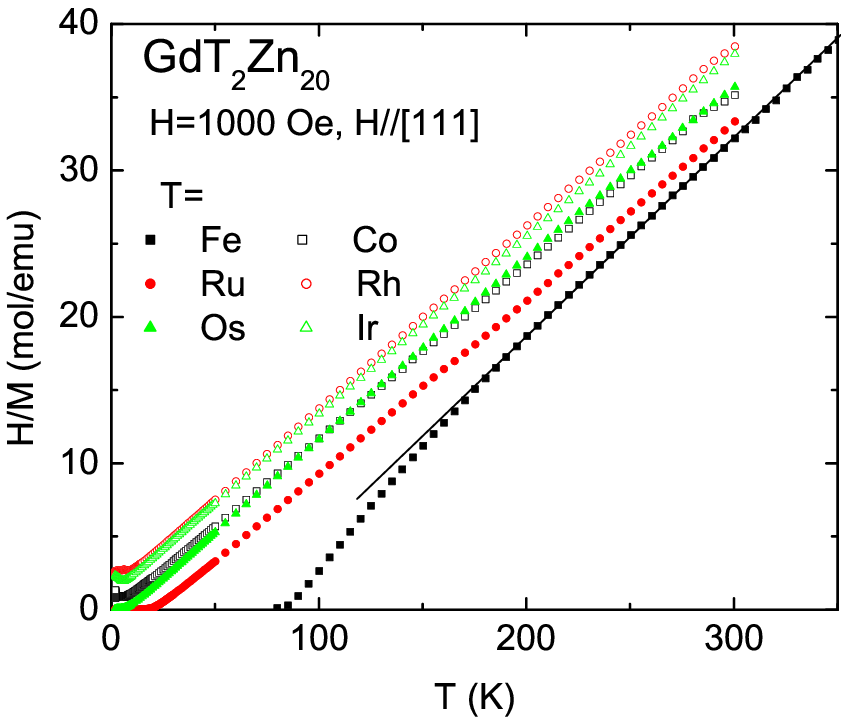}\\
  \caption{(color online) Applied field ($H = 1000$~Oe) divided by the magnetizations of GdT$_2$Zn$_{20}$ as a function of temperature. The solid line represents the high-temperature CW fit for GdFe$_2$Zn$_{20}$.}
  \label{Fig5HMall}
  \end{center}
\end{figure}

Figure \ref{Fig5HMall} presents temperature dependent $H/M$ data for the six Gd based compounds.
For this low magnetic field, $H/M$ approximately equals inverse susceptibility [$1/\chi (T)$] in the paramagnetic state.
Except for GdFe$_2$Zn$_{20}$, the data sets of $1/\chi (T)$ of these compounds are linear and parallel to each other over the whole temperature range of the paramagnetic state, manifesting Curie-Weiss (CW) behavior, $\chi (T)=C/(T-\theta _C)$, where $C$ is Curie constant and $\theta _C$ is paramagnetic Curie temperature.
The same $C$ value is extracted from the parallel lines gives the same effective moments($\mu _{eff}\simeq 8~\mu _B$), close to the value of Hund's rule ground state of Gd$^{3+}$($7.94~\mu_B$), without any apparent contribution from local moments associated with the transition metal.
This is consistent with the low temperature saturated moments, being close to the theoretical value, $\mu _{sat}=7~\mu _B$ (Fig. \ref{Fig4MHall}).
In contrast, $1/\chi (T)$ of GdFe$_2$Zn$_{20}$ obeys a simple CW law only above $\sim 200$~K and evidently deviates from it at lower temperatures (see discussion below).
Nevertheless, its high-temperature CW behavior yields $\mu_{eff}$ close to the others.
The sign of the $\theta _C$ values is consistent with their magnetic ordering type, except for GdCo$_2$Zn$_{20}$, which manifests AFM order but a positive, albeit small, $\theta _C$ (Table \ref{table2}).
This anomalous $\theta _C$ value for GdCo$_2$Zn$_{20}$ leads to a much larger susceptibility near the N\'{e}el temperature, $T_{\mathrm{N}}$, than T = Rh and Ir members (Fig. \ref{Fig3MTall}).

\begin{figure}
  \begin{center}
  \includegraphics[clip, width=0.45\textwidth]{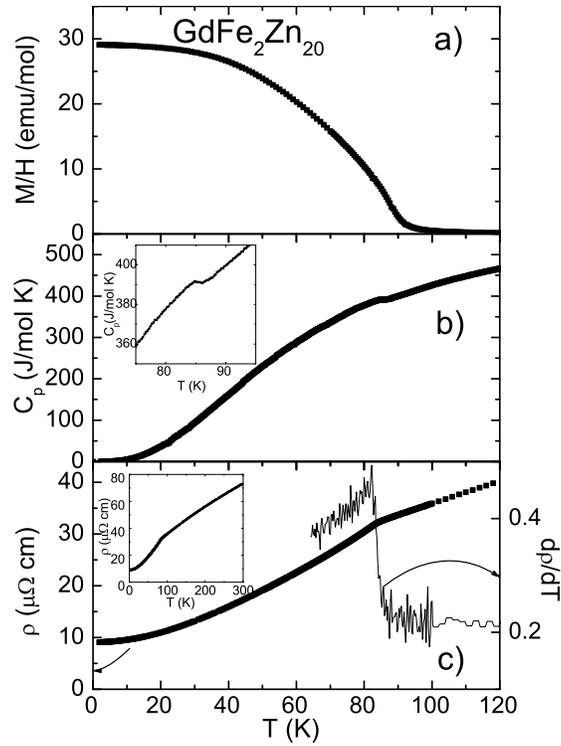}\\
  \caption{(a) Temperature dependent magnetization ($M$) of GdFe$_2$Zn$_{20}$ divided by applied field ($H = 1000$~Oe); (b) specific heat ($C_p$); (c) resistivity ($\rho $)and its derivative respect to temperature ($d\rho /dT$). Inset in (b): detail of $C_p$ data near $T_{\mathrm{C}}$. Inset in (c) $\rho $ over whole temperature range, 2 K~-~300 K.}
  \label{Fig6GdFetotal}
  \end{center}
\end{figure}

\begin{figure}
  \begin{center}
  \includegraphics[clip, width=0.45\textwidth]{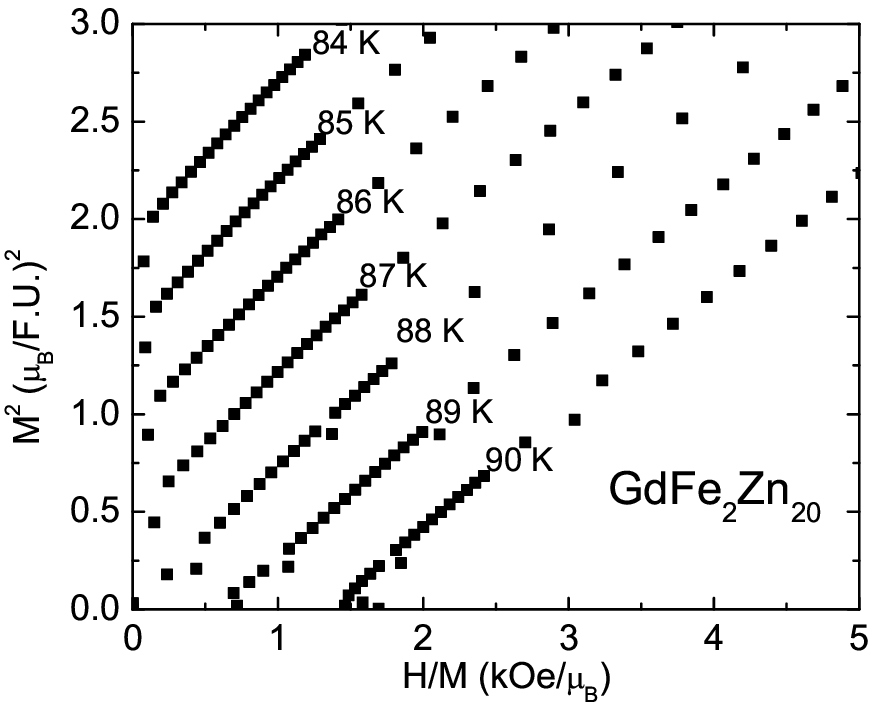}\\
  \caption{Arrott plot for GdFe$_2$Zn$_{20}$.}
  \label{Fig7GdFearrott}
  \end{center}
\end{figure}

GdFe$_2$Zn$_{20}$ is the most conspicuously anomalous in its behavior.
Figure \ref{Fig6GdFetotal} presents a blow up of the low field $M/H$ data as well as the results of measurements of temperature dependent specific heat ($C_p$) and electrical resistivity ($\rho$) in zero applied magnetic field.
The specific heat data manifest a clear anomaly at $T_{\mathrm{C}}= 85\pm 1$~K [inset of Fig. \ref{Fig6GdFetotal}(b)].
The resistivity data, although collected from a sample from different batch, show a clear break in slope (or maximum in $d\rho /dT$) at $T_{\mathrm{C}}= 84\pm 2$~K.
Determination of the ordering temperature from magnetization data requires a more detailed analysis.
Figure \ref{Fig7GdFearrott} presents a plot of $M^2$ versus $H/M$ (an Arrott plot)\cite{arrott_criterion_1957} from data collected on the same batch of sample used for $C_p$ in the vicinity of $T_{\mathrm{C}}$.
The isotherm that most closely goes linearly through the origin is the one closest to $T_{\mathrm{C}}$, giving a value 88 K.
All of these measurements are consistent with transition temperature near 86 K.
It should be noted though, that $T_{\mathrm{C}}$ values for different batch of samples can vary by as much as $\pm 3$~K \cite{jia_nearly_2007}, even though the single-crystal X-ray measurements do not suggest evident crystallographic difference.

\begin{figure}
  \begin{center}
  \includegraphics[clip, width=0.45\textwidth]{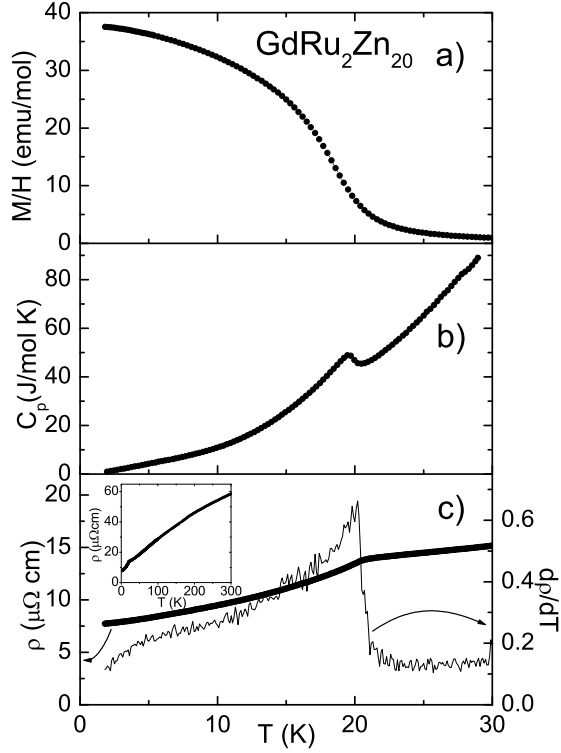}\\
  \caption{(a) Temperature dependent $M/H$ for GdRu$_2$Zn$_{20}$ ($H = 1000$~Oe); (b) $C_p$; (c) $\rho$  and $d\rho /dT$. Inset in (c): $\rho $ over whole temperature range.}
  \label{Fig8GdRutotal}
  \end{center}
\end{figure}

\begin{figure}
  \begin{center}
  \includegraphics[clip, width=0.45\textwidth]{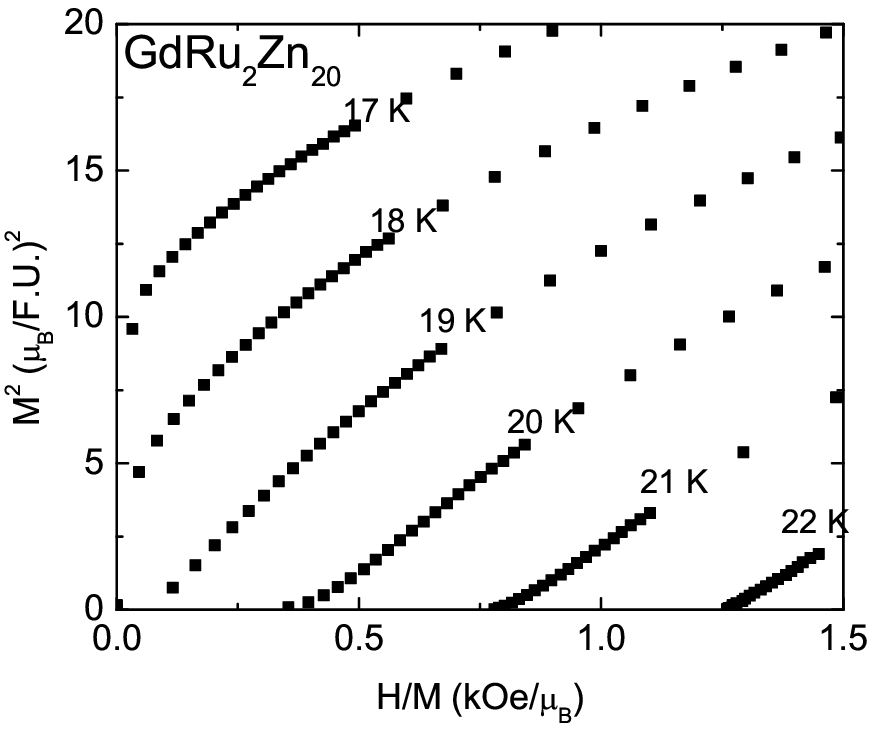}\\
  \caption{Arrott plot for GdRu$_2$Zn$_{20}$.}
  \label{Fig9GdRuarrott}
  \end{center}
\end{figure}

GdRu$_2$Zn$_{20}$ also manifests a relatively high FM ordering temperature.
Figures \ref{Fig8GdRutotal}(b, c) present temperature dependent specific heat and electrical resistivity measurements on GdRu$_2$Zn$_{20}$ in zero applied magnetic fields, both of which show clear evidence of ordering with $T_{\mathrm{C}}=20\pm 1$~K.
Figure \ref{Fig9GdRuarrott} shows that, similar to GdFe$_2$Zn$_{20}$, the $T_{\mathrm{C}}$ of GdRu$_2$Zn$_{20}$ can be inferred from an Arrott plot analysis.
These measurements were performed on samples from the same batch and the different methods for determining $T_{\mathrm{C}}$ agree to within $\pm 1$~K difference.

\begin{figure}
  \begin{center}
  \includegraphics[clip, width=0.45\textwidth]{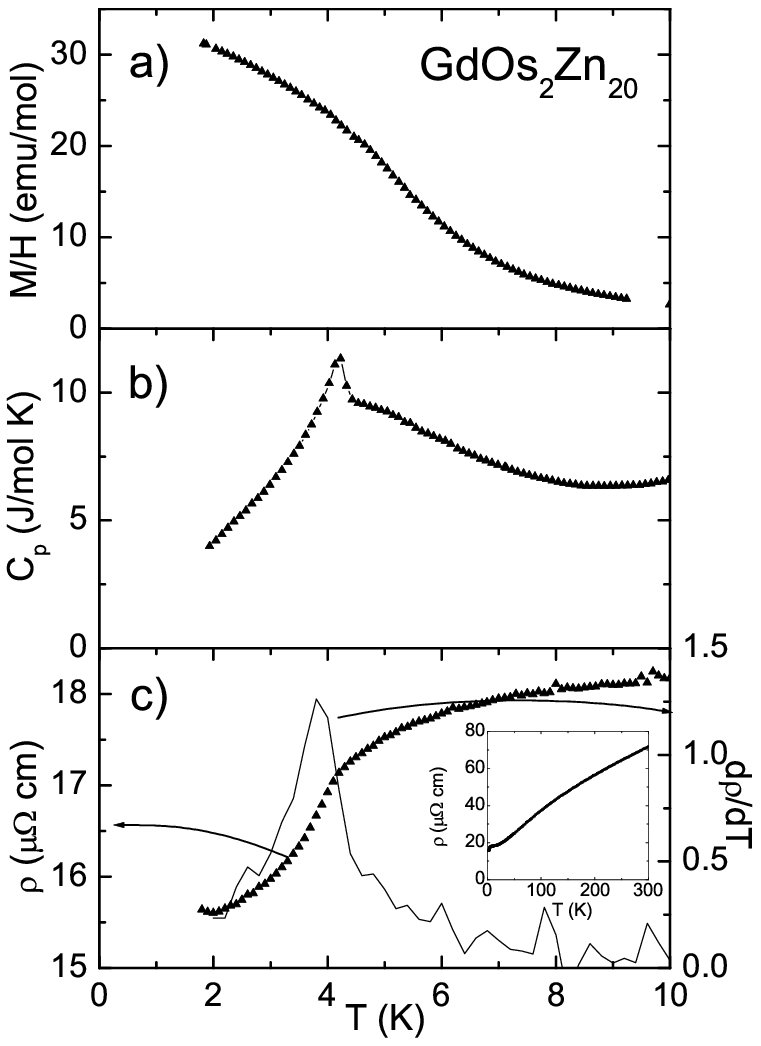}\\
  \caption{(a) Temperature dependent $M/H$ for GdOs$_2$Zn$_{20}$ ($H = 1000$ Oe); (b) $C_p$; (c) $\rho $ and $d\rho /dT$. Inset in (c): $\rho $ over whole temperature range.}
  \label{Fig10GdOstotal}
  \end{center}
\end{figure}

\begin{figure}
  \begin{center}
  \includegraphics[clip, width=0.45\textwidth]{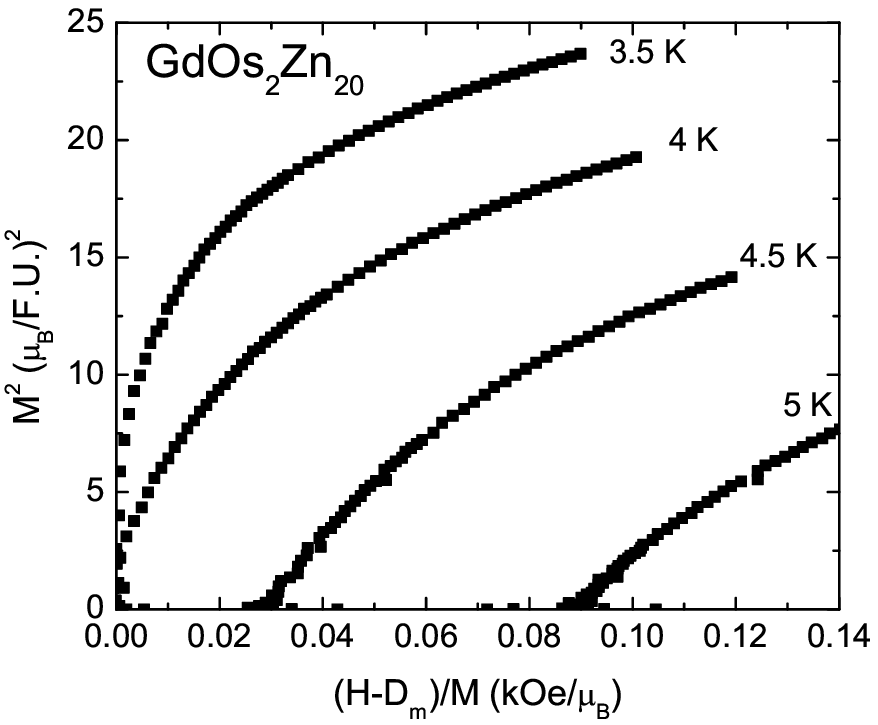}\\
  \caption{Arrott plot for GdOs$_2$Zn$_{20}$. The demagnetizing field $D_m$ can not be ignored for this low $T_{\mathrm{C}}$, and was estimated from the geometric factor of the sample ($D\sim 0.03$).}
  \label{Fig11GdOsarrott}
  \end{center}
\end{figure}

GdOs$_2$Zn$_{20}$ appears to order ferromagnetically at a $T_{\mathrm{C}}$ value as low as the N\'{e}el temperatures found for the Co column members of the GdT$_2$Zn$_{20}$ family (see below).
As shown in Fig. \ref{Fig10GdOstotal}(b) and (c), the specific heat and resistivity data manifest features consistent with a magnetic phase transition near 4 K.
However, the $C_p$ data, with a broad shoulder above this temperature, does not manifest a standard $\lambda $-type of feature and may indicate a distribution of $T_{\mathrm{C}}$ values or multiple transitions.  
The Arrott plot for GdOs$_2$Zn$_{20}$, although having non-linear, isothermal curves, is also consistent with a FM transition between 4 K and 4.5 K (Fig. \ref{Fig11GdOsarrott}).
Such a non-linear feature in the isothermal curves is also found in ref.\cite{brommer_strongly_1990, yeung_arrott-plot_1986}, and may be associated with complex magnetic phenomenon in the critical region, rather than one simple, clearly defined, Landau type, 2nd order phase transition.

\begin{figure}
  \begin{center}
  \includegraphics[clip, width=0.45\textwidth]{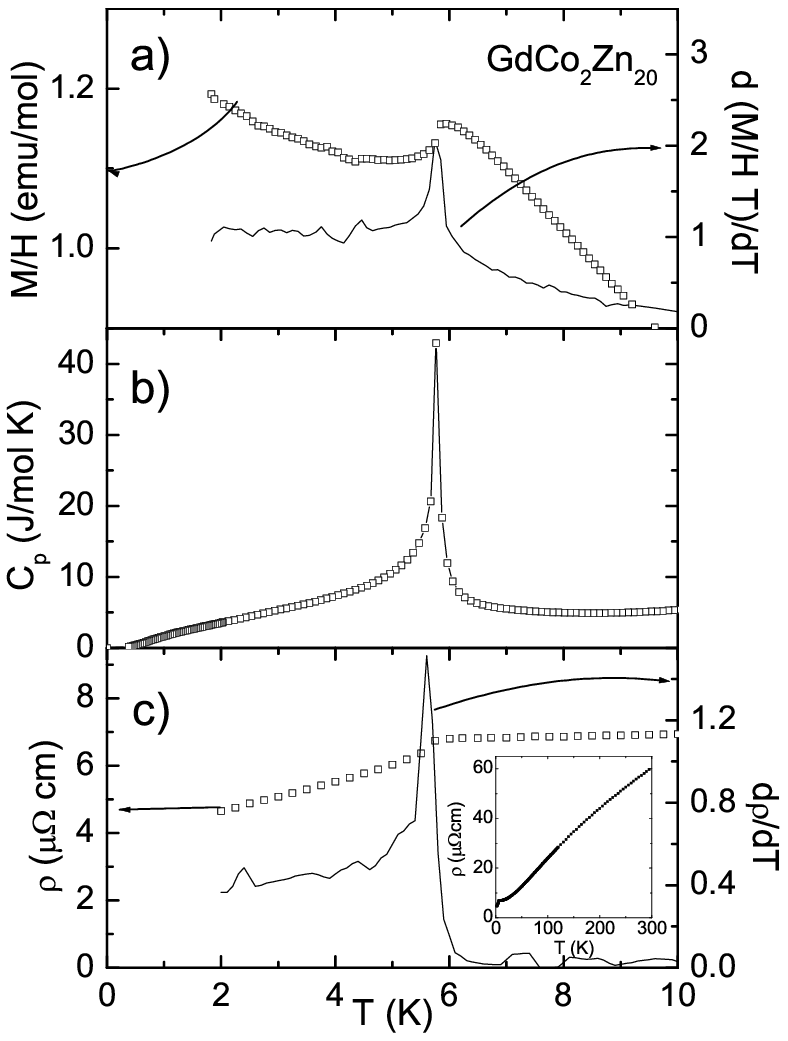}\\
  \caption{(a) Temperature dependent susceptibility ($\chi $) and $d(\chi T)/dT$ of GdCo$_2$Zn$_{20}$; (b) $C_p$; (c) $\rho $ and $d\rho /dT$. Inset in (c): $\rho $ over whole temperature range.}
  \label{Fig12GdCo}
  \end{center}
\end{figure}

\begin{figure}
  \begin{center}
  \includegraphics[clip, width=0.45\textwidth]{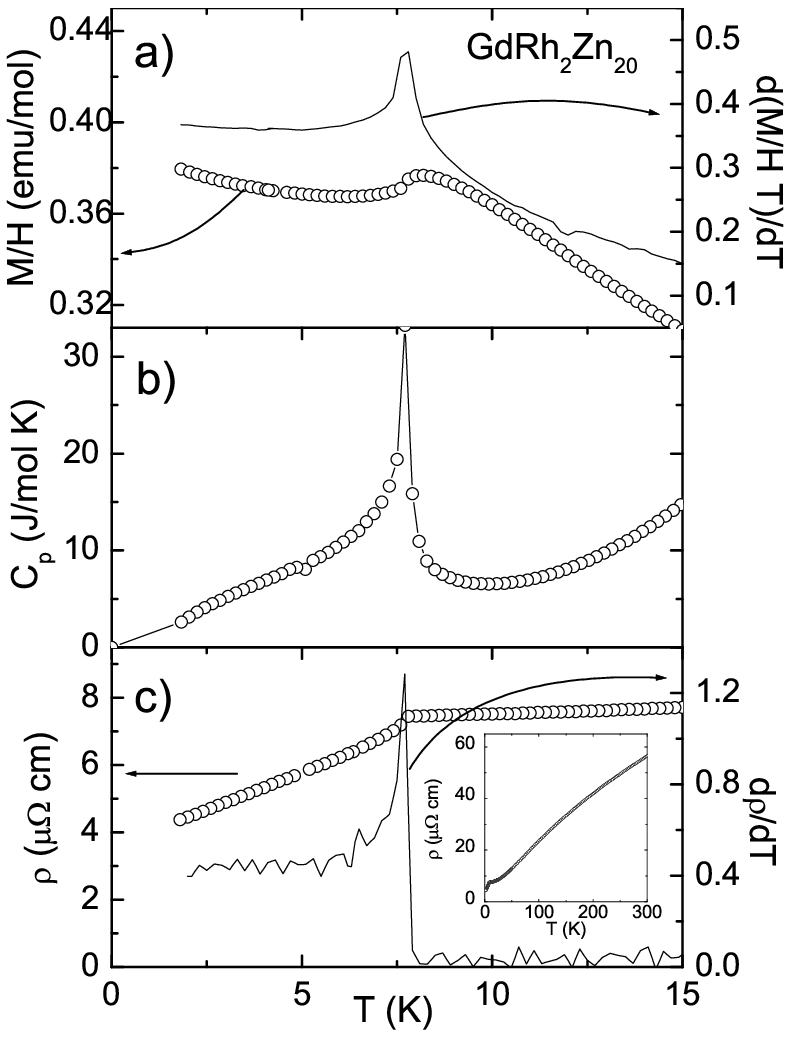}\\
  \caption{(a) Temperature dependent $\chi $ and $d(\chi T)/dT$ of GdRh$_2$Zn$_{20}$; (b) $C_p$; (c) $\rho $ and $d\rho /dT$. Inset in (c): $\rho $ over whole temperature range.}
  \label{Fig13GdRh}
  \end{center}
\end{figure}

\begin{figure}
  \begin{center}
  \includegraphics[clip, width=0.45\textwidth]{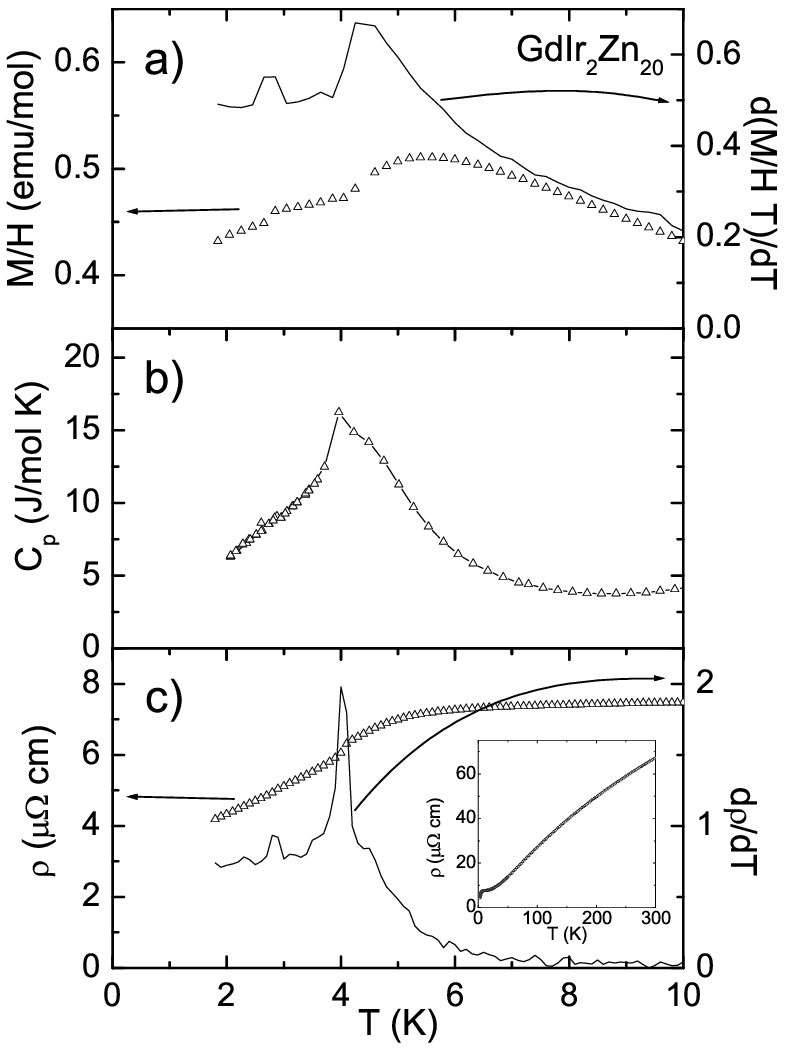}\\
  \caption{(a) Temperature dependent $\chi $ and $d(\chi T)/dT$ of GdIr$_2$Zn$_{20}$; (b) $C_p$; (c) $\rho $ and $d\rho /dT$. Inset in (c): $\rho $ over whole temperature range.}
  \label{Fig14GdIr}
  \end{center}
\end{figure}

In contrast to the Fe column compounds, the Co column compounds all appear to order antiferromagnetically with the values of $T_{\mathrm{N}}$ between 4 and 7 K.
Figures \ref{Fig12GdCo}, \ref{Fig13GdRh} and \ref{Fig14GdIr} present the low temperature magnetic susceptibility, specific heat and electrical resistivity data for GdCo$_2$Zn$_{20}$, GdRh$_2$Zn$_{20}$ and GdIr$_2$Zn$_{20}$ respectively.
In addition to these data, $d(\chi (T) T)/dT$~\cite{fisher1962} and $d\rho /dT$~\cite{fisher_resistive_1968} have been added to the susceptibility and resistivity plots respectively.
GdCo$_2$Zn$_{20}$ and GdRh$_2$Zn$_{20}$ manifest clear $\lambda$-type anomalies in their temperature dependent specific heat, with similar features appearing in their $d\rho /dT$ and $d(\chi (T) T)/dT$ data.
From these thermodynamic and transport data we infer $T_{\mathrm{N}}$ of 5.7 K and 7.6 K for GdCo$_2$Zn$_{20}$ and GdRh$_2$Zn$_{20}$ respectively.
GdIr$_2$Zn$_{20}$ shows a somewhat broader feature at $T_{\mathrm{N}}=4$~K and there may be a lower temperature transition near 2 K indicated in the magnetization data, although this is not clearly supported by corresponding features in either specific heat or resistivity data.
A summary of the thermodynamic and transport measurements on the six GdT$_2$Zn$_{20}$ compounds is presented in Table \ref{table2}.

\begin{table}
\caption{\label{table2} Residual resistivity ratio, $RRR = \frac{R(300K)}{R(2K)}$; paramagnetic Curie temperature, $\theta _C$ and effective moment, $\mu _{eff}$ (from the CW fit of $\chi (T)$ from 100 K to 300 K, except for GdFe$_2$Zn$_{20}$; see text for details); magnetic ordering temperature, $T_{mag}$; and  saturated moment at 55 kOe along [111] direction, $\mu _{sat}$ on GdT$_2$Zn$_{20}$ compounds (T = Fe, Ru, Os, Co, Rh, Ir).}
\begin{ruledtabular}
\begin{tabular}{ccccccc}
T & Fe & Ru & Os & Co & Rh & Ir \\
\hline
$RRR$ & 8.1 & 7.6 & 5 & 12.8 & 12.8 & 15.7\\
$\theta _C$, K & 46 & 23 & 3 & 3 & -10 & -8\\
$\mu _{eff}$, $\mu _{B}$ & 7.9 & 8.2 & 8.1 & 8.2 & 8.0 & 8.1\\
$T_{mag}$, K & 86 & 20 & 4.2 & 5.7 & 7.7 & 4.2, 2.4\footnote{two magnetic transitions were found}\\
$\mu _{sat}$, $\mu _{B}$ & 6.5 & 7.25 & 6.9 & 7.3 & & \\
\end{tabular}
\end{ruledtabular}
\end{table}

\begin{figure}
  \begin{center}
  \includegraphics[clip, width=0.45\textwidth]{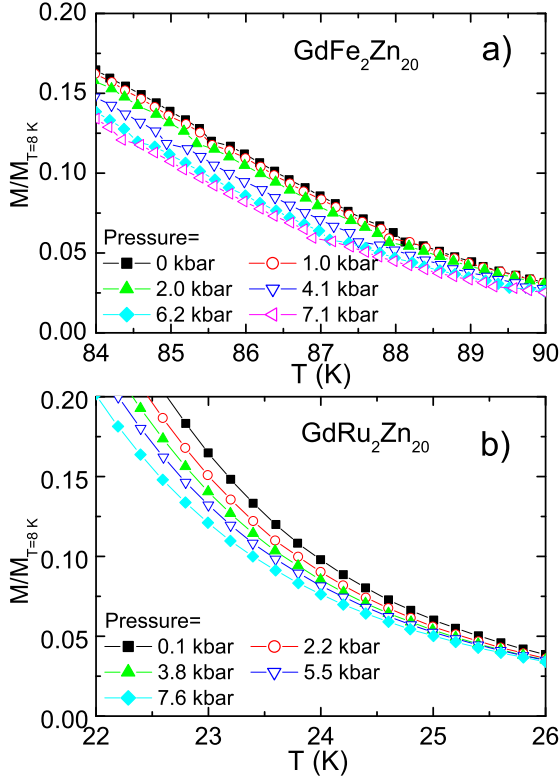}\\
  \caption{(color online) Magnetization of (a): GdFe$_2$Zn$_{20}$ and (b): GdRu$_2$Zn$_{20}$ in applied field ($H = 1000$ Oe) under different hydrostatic pressure.}
  \label{Fig15pressure}
  \end{center}
\end{figure}

\begin{figure}
  \begin{center}
  \includegraphics[clip, width=0.45\textwidth]{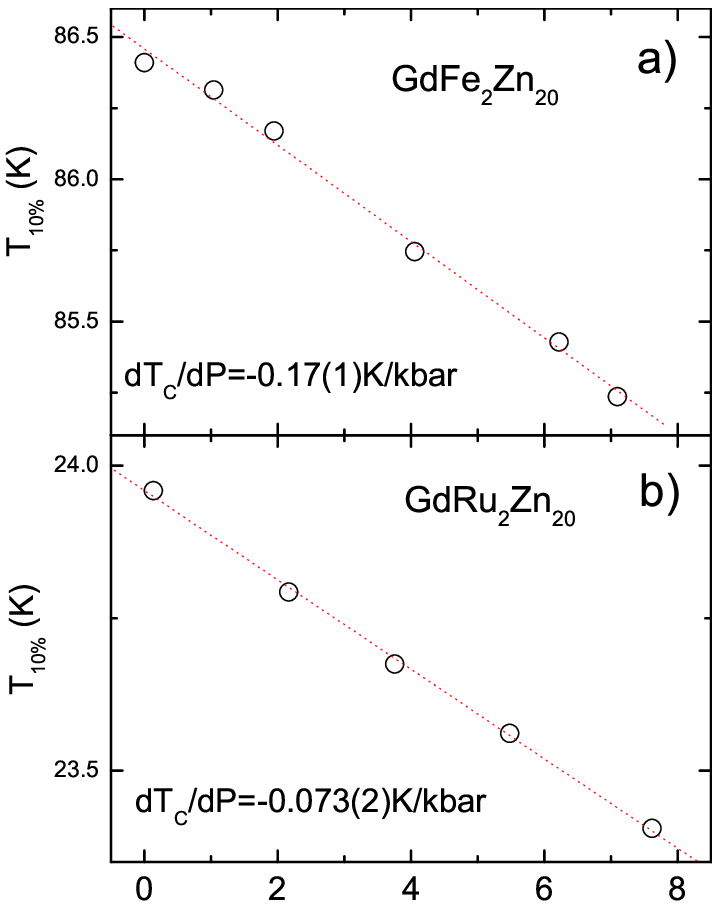}\\
  \caption{Pressure dependent $T_{10\%}$ (inferred as $T_{\mathrm{C}}$) of GdFe$_2$Zn$_{20}$ and GdRu$_2$Zn$_{20}$. The dash lines are the linear fits of the data.}
  \label{Fig16pressureTc}
  \end{center}
\end{figure}

A logical question that comes to mind when comparing $T_{\mathrm{C}}$ for the Fe column members with the lattice parameter data shown in Fig. \ref{Fig2lattice} is whether the drop in $T_{\mathrm{C}}$ as the transition metal moves down the column is associated with a steric effect.
This can be addressed experimentally by measurements of $T_{\mathrm{C}}$ under hydrostatic pressure.
Figure \ref{Fig15pressure} presents low field magnetization for GdFe$_2$Zn$_{20}$ and GdRu$_2$Zn$_{20}$ under pressures up to 7 kilobar.
The application of pressure suppresses the ferromagnetically ordered state in both compounds and the pressure dependence of $T_{10\%}$ (the temperature where the magnetization equals 10\% of maximum magnetization and used as a caliper of $T_{\mathrm{C}}$) of each compound is plotted in Fig. \ref{Fig16pressureTc}.
The fact that both compounds manifest a suppression of $T_{\mathrm{C}}$ with increasing pressure indicates that the difference between GdFe$_2$Zn$_{20}$ and GdRu$_2$Zn$_{20}$ is not primarily a steric one.
Approximating the bulk modulus of these compounds to be a generic 1Mbar, one can estimate that GdRu$_2$Zn$_{20}$ under 10 kbar hydrostatic pressure will have its lattice parameter reduced by 0.03 \AA (25\% of the difference between the lattice parameter of GdFe$_2$Zn$_{20}$ and GdCo$_2$Zn$_{20}$).
If the cause of the $T_{\mathrm{C}}$ suppression was purely steric, such a change in lattice parameter should (at the very least) result in a dramatic increase in the $T_{\mathrm{C}}$ values of GdRu$_2$Zn$_{20}$ rather than the gradual suppression observed.

\subsection{\label{sec:C}YT$_2$Zn$_{20}$(T = Fe, Co, Ru, Rh, Os and Ir)}

\begin{figure}
  \begin{center}
  \includegraphics[clip, width=0.45\textwidth]{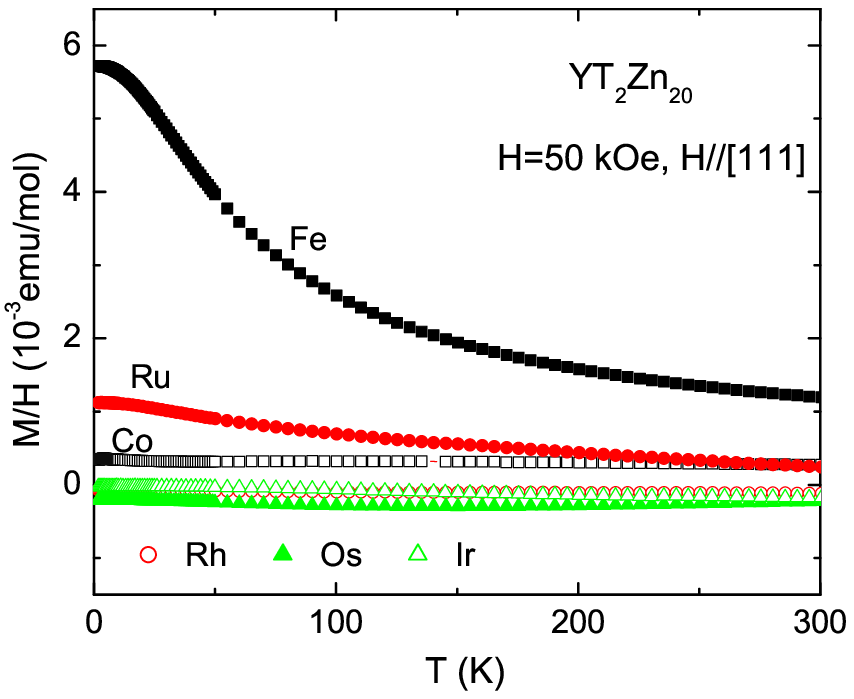}\\
  \caption{(color online) Temperature dependent magnetization of YT$_2$Zn$_{20}$ under applied field $H = 50$ kOe.}
  \label{Fig17YMT}
  \end{center}
\end{figure}

\begin{figure}
  \begin{center}
  \includegraphics[clip, width=0.45\textwidth]{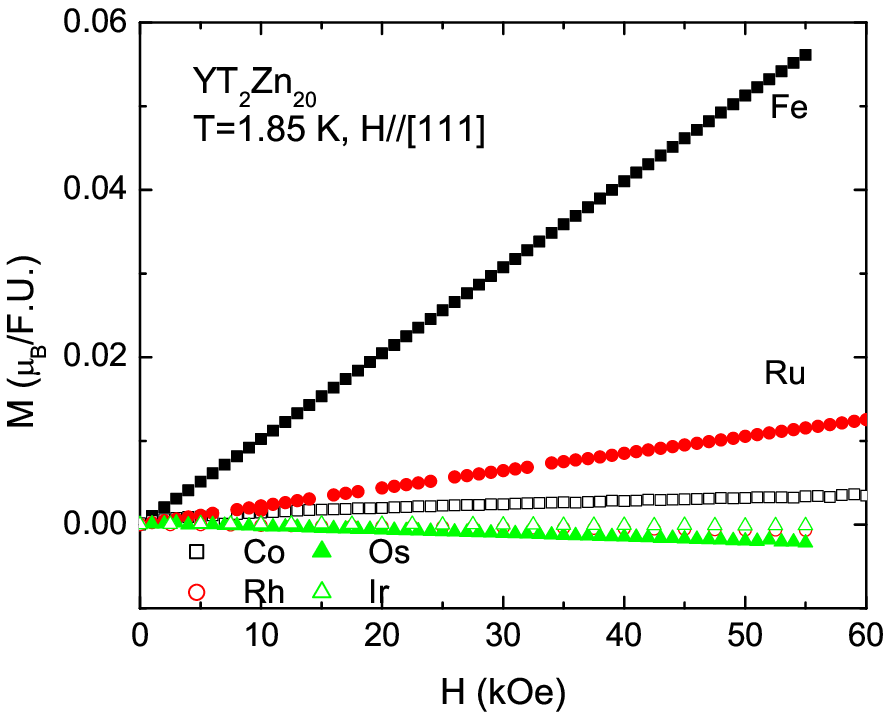}\\
  \caption{(color online) Field dependent magnetization of YT$_2$Zn$_{20}$ at 1.85~K.}
  \label{Fig18YMH}
  \end{center}
\end{figure}

In order to better understand the behavior of GdFe$_2$Zn$_{20}$ and GdRu$_2$Zn$_{20}$ with respect to the rest of the GdT$_2$Zn$_{20}$ compounds, it is useful to examine the properties of the nonmagnetic analogues: the YT$_2$Zn$_{20}$ compounds.
The temperature dependent magnetization data (divided by applied field) and the low temperature magnetization isotherms for these six compounds are presented in Fig. \ref{Fig17YMT} and Fig. \ref{Fig18YMH}, respectively.
YFe$_2$Zn$_{20}$ and YRu$_2$Zn$_{20}$ have a greatly and intermediately enhanced paramagnetic signals respectively, whereas the rest of the materials manifest the ordinary weak, either paramagnetic or diamagnetic, response, anticipated for non-moment bearing intermetallic compounds.

\begin{figure}
  \begin{center}
  \includegraphics[clip, width=0.45\textwidth]{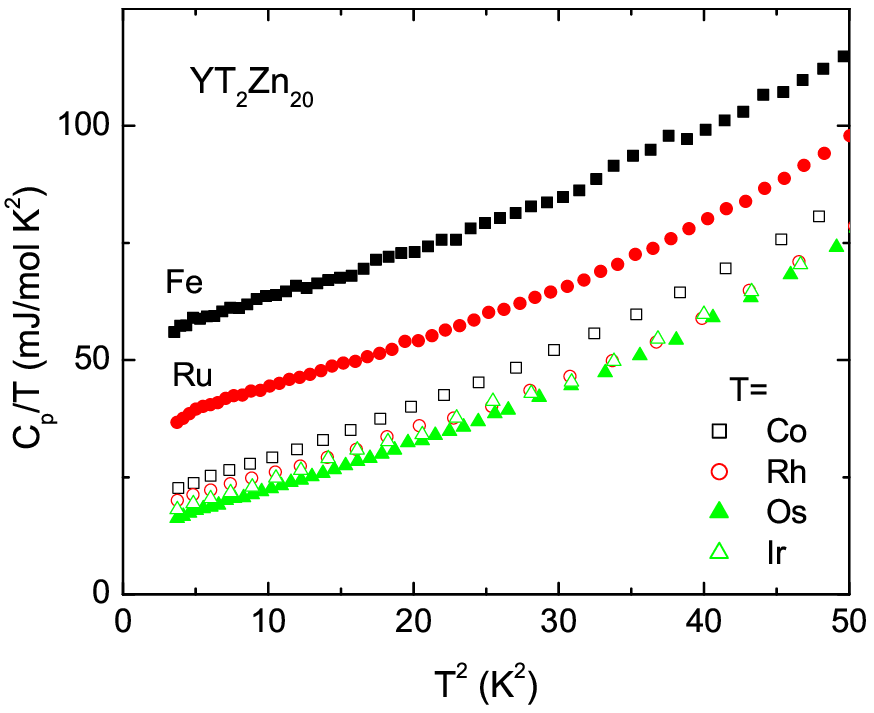}\\
  \caption{(color online) Low temperature specific heat of YT$_2$Zn$_{20}$.}
  \label{Fig19YCp}
  \end{center}
\end{figure}

Measurements of low temperature specific heat (plotted as $C_p/T$ versus $T^2$ in Fig. \ref{Fig19YCp}) also indicate a clear difference between YFe$_2$Zn$_{20}$, YRu$_2$Zn$_{20}$ and the other members of the YT$_2$Zn$_{20}$ series: enhanced values of the electronic specific heat being found for T = Fe and Ru.
As previously reported\cite{jia_nearly_2007}, YFe$_2$Zn$_{20}$ can be thought of as being close to the Stoner limit. The simplest way to see this is to recall that, in this limit, whereas the Pauli paramagnetism is enhanced by a factor $(1-Z)^{-1}$, the electronic specific heat is not\cite{ziman_principles_1979}. This means that the term $Z$ in the enhancement factor can then be inferred from the experimentally determined, low temperature values of $\gamma _0$ and $\chi _0$. In common units

\begin{equation}
Z=1-1.37\times 10^{-2}\frac{\gamma _0(J/molK^2)}{\chi _{0-dia}(emu/mol)}
\label{eqn:2}
\end{equation}
where $\chi _{0-dia}$ equals $\chi _0$ with the core diamagnetism subtracted.

Giving the core diamagnetism values($-2.3\times 10^{-4}$emu/mol for YFe$_2$Zn$_{20}$ and YCo$_2$Zn$_{20}$, $-2.5\times 10^{-4}$emu/mol for YRu$_2$Zn$_{20}$ and YRh$_2$Zn$_{20}$, and $-2.9\times 10^{-4}$emu/mol for YOs$_2$Zn$_{20}$ and YIr$_2$Zn$_{20}$)\cite{mulay_theory_1976}, $Z$ can be inferred to be 0.88 and 0.67 for YFe$_2$Zn$_{20}$ and YRu$_2$Zn$_{20}$ respectively (Table \ref{table3}).
For reference, this can be compared to $Z = 0.83$ and $0.57$ for elemental Pd and Pt respectively\cite{pdptZ}, which are thought to be canonical examples of NFFL. These enhanced $Z$ values indicate that YRu$_2$Zn$_{20}$, and particular YFe$_2$Zn$_{20}$ are extremely close to the Stoner limit ($Z = 1$).
In contrast, the $Z$ values of the rest of the members are less than $0.5$, which is comparable with the estimated value of the canonical example of `normal metal', Cu, $Z=0.29$ \cite{CuAgAu}.
It is worth to notice that, during the estimation of the $Z$ values, the contribution from the Landau diamagnetism is ignored.
Inversely proportional to the square of the effective mass of the conduction electrons\cite{ashcroft_solid_1976}, the Landau diamagnetic contribution becomes more significant for those members which have smaller $\gamma _0$ values.
Thus, based on the thermodynamic measurements, the Pauli susceptibility values, even after the core diamagnetism correction, are still under-estimated.
Due to this uncertainty, the Pauli susceptibility values after the core diamagnetism correction for YOs$_2$Zn$_{20}$ and YRh$_2$Zn$_{20}$, albeit positive, are still less than the un-enhanced values ($Z=0$) corresponding to their $\gamma _0$. 

\begin{table}
\caption{\label{table3} Low temperature susceptibility, $\chi _0$; and the values after core diamagnetism correction, $\chi _{0-dia}$; linear coefficient of the specific heat, $\gamma _0$; and the Stoner enhancement factor, $Z$ on YT$_2$Zn$_{20}$ compounds (T = Fe, Ru, Os, Co, Rh, Ir). }
\begin{ruledtabular}
\begin{tabular}{ccccccc}
T & Fe & Ru & Os & Co & Rh & Ir \\
\hline
$\chi _0$\footnote{Taken as $\frac{M(50~kOe)-M(30~kOe)}{20~kOe}$,in unit $10^{-3}$emu/mol}, & 5.73 & 1.14 & -0.256 & 0.212 & -0.076 & -0.034\\
$\chi _{0-dia}$, & 5.96 & 1.39 & 0.034 & 0.442 & 0.174 & 0.256\\
$\gamma _0$\footnote{in unit mJ/molK$^2$} & 53 & 34 & 12.4 & 18.3 & 16.4 & 14.1\\
$\theta _D$, K & 123 & 124 & 125 & 121 & 127 & 124\\
$Z$\footnote{Eqn. \ref{eqn:2} is invalid for T = Os and Rh; see text} & 0.88 & 0.67 & - & 0.43 & - & 0.24\\
\end{tabular}
\end{ruledtabular}
\end{table}

\subsection{\label{sec:D}Electronic Structure}

\begin{figure}
  \begin{center}
  \includegraphics[clip,angle=270, width=0.45\textwidth]{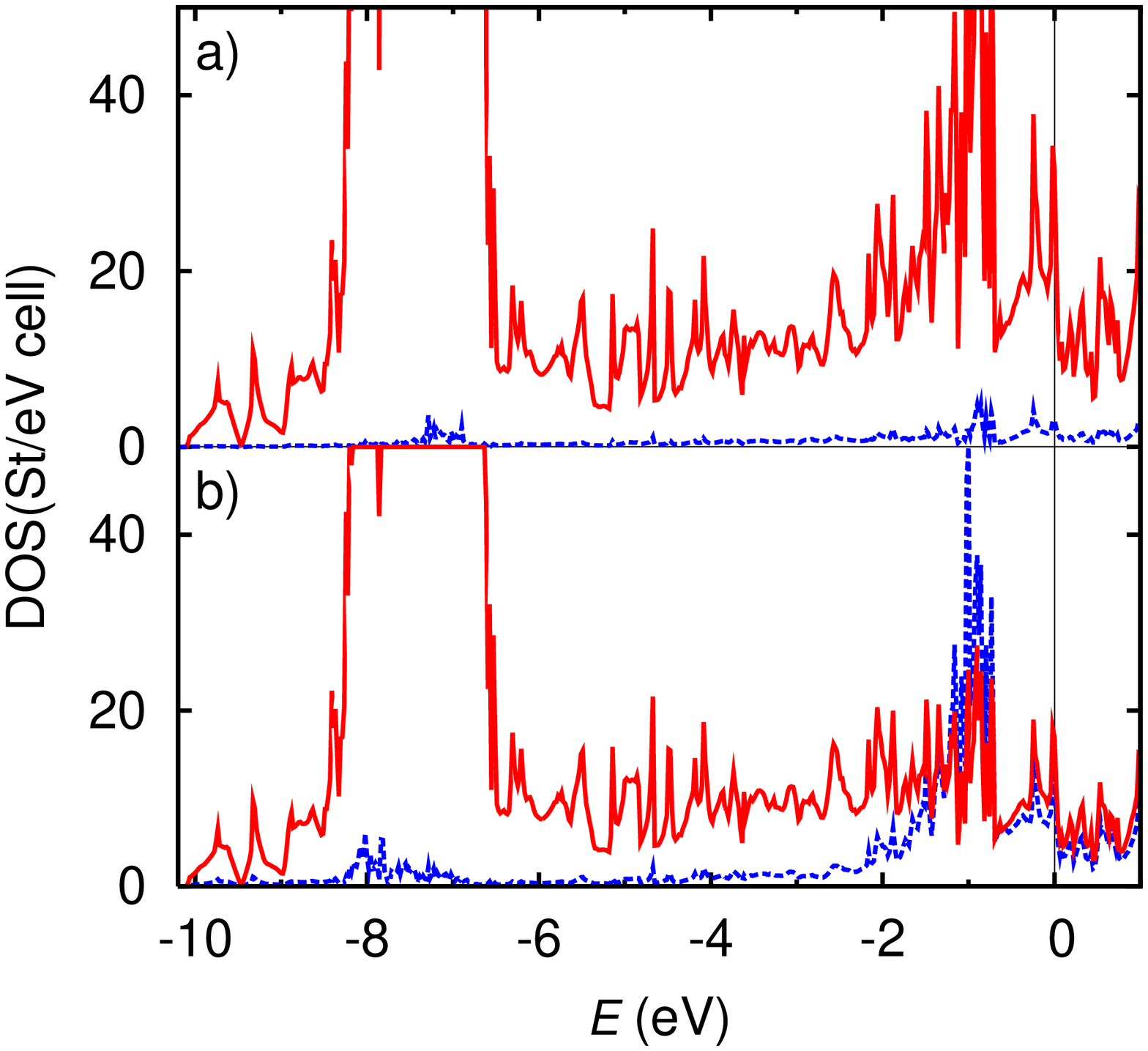}\\
  \caption{(color online) The DOS of YFe$_2$Zn$_{20}$ (in St/eV cell) and partial DOS 
(in St/eV cell). $E_F$ corresponds to zero energy. The red color solid line on (a) corresponds to total DOS and blue dashed - to Y atoms PDOS. The red color solid 
line on (b) corresponds to PDOS of Zn and blue dashed - to Fe atoms PDOS.}
  \label{Fig20DOSYFe}
  \end{center}
\end{figure}

Band structure calculations, performed on the representative, non-local moment members, YT$_2$Zn$_{20}$ (T = Fe, Co and Ru), as well as their local moment analogues, GdT$_2$Zn$_{20}$, can help us to understand their diverse magnetic properties further.
Figure \ref{Fig20DOSYFe} shows the result of the total and partial density of states (DOS) for each element for YFe$_2$Zn$_{20}$.
At the Fermi level, $E_F$, the total DOS manifests a sharp peak, leading to the relatively large DOS at Fermi level ($N(E_F)$, see Table \ref{table4}), and therefore large band contributed electronic specific heat, $\gamma _{band}=30.6$ mJ/mol K$^2$.
This result is consistent with the experimentally measured electronic specific heat, $\gamma _0$ with a large mass enhanced factor, $\lambda = 0.73$, if one assumes $\gamma _0=(1+\lambda )\gamma _{band}$.
The peak-shape DOS at $E_F$ is not unusual for the NFFL systems: similar calculation results have been obtained for Pd\cite{shimizu_pdband_1963}, YCo$_2$\cite{tanaka_mass_1998} and Ni$_3$Ga\cite{hayden_electronic_1986} by using similar techniques.
The large peak at about $-7$~eV corresponds to totally filled $d$-states of Zn atoms. 
Figure \ref{Fig20DOSYFe} also shows significant contribution of Zn atoms' electronic states to the total DOS in the whole energy spectrum, whereas the Fe atoms' electronic states are mostly localized in the vicinity of $E_F$, although they are dilute in this compound (1/10 of Zn).
Table \ref{table4} shows that the partial DOS of Fe at $E_F$ is in between the values for elemental Pd and Fe (before band splitting), the canonical examples of nearly ferromagnet and `strong' ferromagnet systems.
This result indicates that YFe$_2$Zn$_{20}$ indeed may be even closer the Stoner criterion than Pd.
The total DOS at $E_F$ mainly corresponds to the hybridization of the $3d$-band of Fe and $p$-band of Zn; the $4d$-band of Y, although hybridized with the other two, contributes significantly less (Fig.\ref{Fig20DOSYFe}).

\begin{figure}
  \begin{center}
  \includegraphics[clip, angle=270, width=0.45\textwidth]{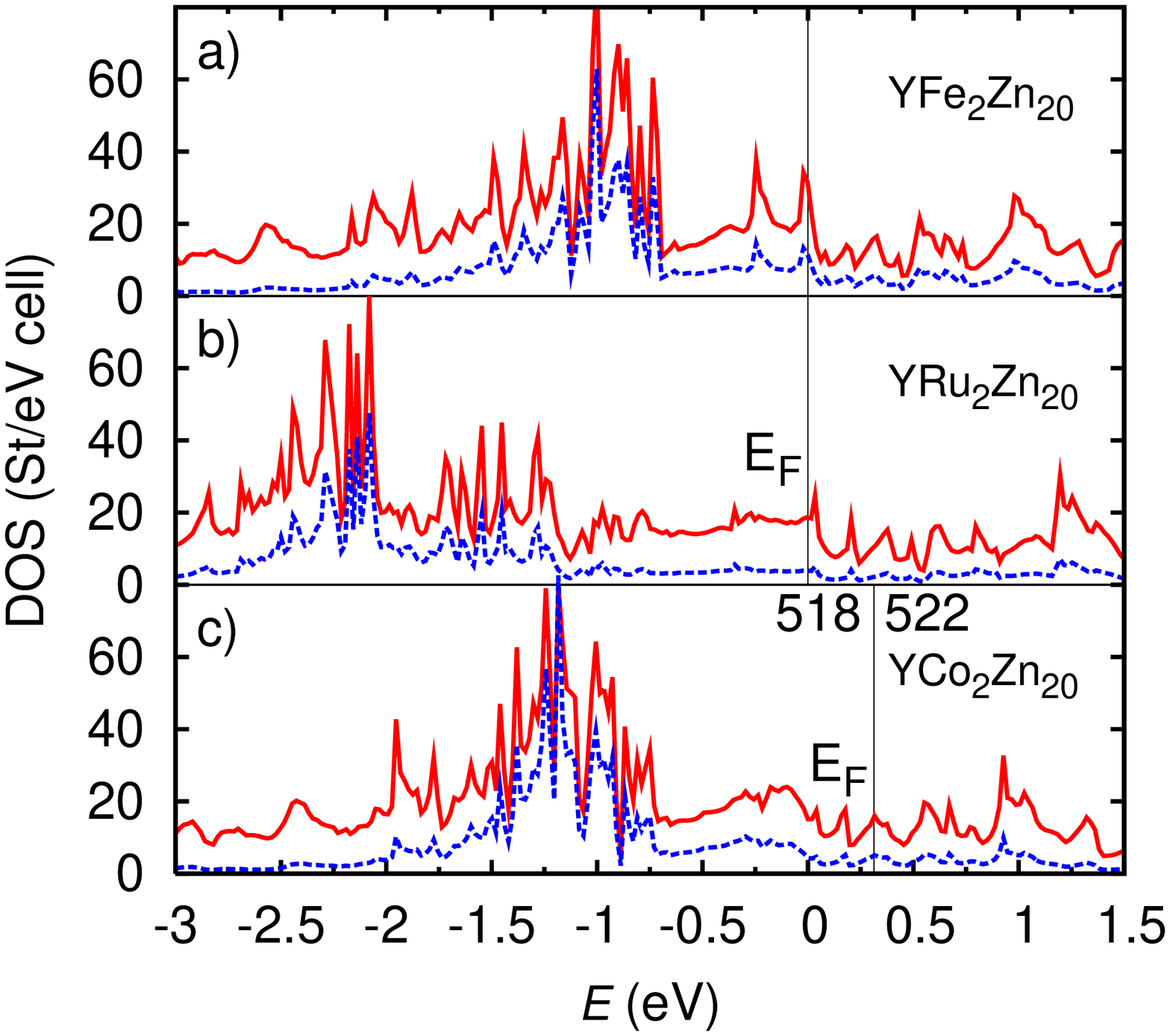}\\
  \caption{(color online) The DOS of YFe$_2$Zn$_{20}$ (a), YRu$_2$Zn$_{20}$ (b) and 
YCo$_2$Zn$_{20}$ (c) near $E_F$ (in St/eV cell) shown by red solid line and PDOS of Fe, Ru and 
Co atoms (blue dashed line) (in St/eV cell). $E_F$ is shown by vertical lines. 518 and 522 corresponds 
to number of valence electrons in the unit cell calculated in the rigid band approximation from the
DOS of YFe$_2$Zn$_{20}$.}
  \label{Fig21DOSYall}
  \end{center}
\end{figure}

The dominant effect of the $d$-band filling on the magnetic properties of YT$_2$Zn$_{20}$, manifests itself clearer if one compares the electronic structure of the three YT$_2$Zn$_{20}$ compounds: T= Fe, Co and Ru (Fig. \ref{Fig21DOSYall}).
In Fig. \ref{Fig21DOSYall}, the total and Co-partial DOS of YCo$_2$Zn$_{20}$ represents behavior similar to the YFe$_2$Zn$_{20}$ analogues, whereas $E_F$ is shifted 0.3 eV higher by adding 2 more valence electrons per formula unit.
This similarity indicates that the difference in the electronic structure of YFe$_2$Zn$_{20}$ and YCo$_2$Zn$_{20}$ can be considered in terms of the rigid band approximation.
On the other hand, the electronic structure of YRu$_2$Zn$_{20}$ has the same Fermi level position as YFe$_2$Zn$_{20}$ because of the same valence electron filling.
However, its total, and partial-Ru, DOS are lower than those for YFe$_2$Zn$_{20}$.
This difference is not unexpected, since the $4d$ band is usually broader than the $3d$ band in the electronic structure of intermetallics.
Calculated $N(E_F)$ of YCo$_2$Zn$_{20}$ is half of the value of YFe$_2$Zn$_{20}$, whereas the value of YRu$_2$Zn$_{20}$ is slightly larger than YCo$_2$Zn$_{20}$ (Table \ref{table4}). 

\begin{figure}
  \begin{center}
  \includegraphics[clip, width=0.45\textwidth]{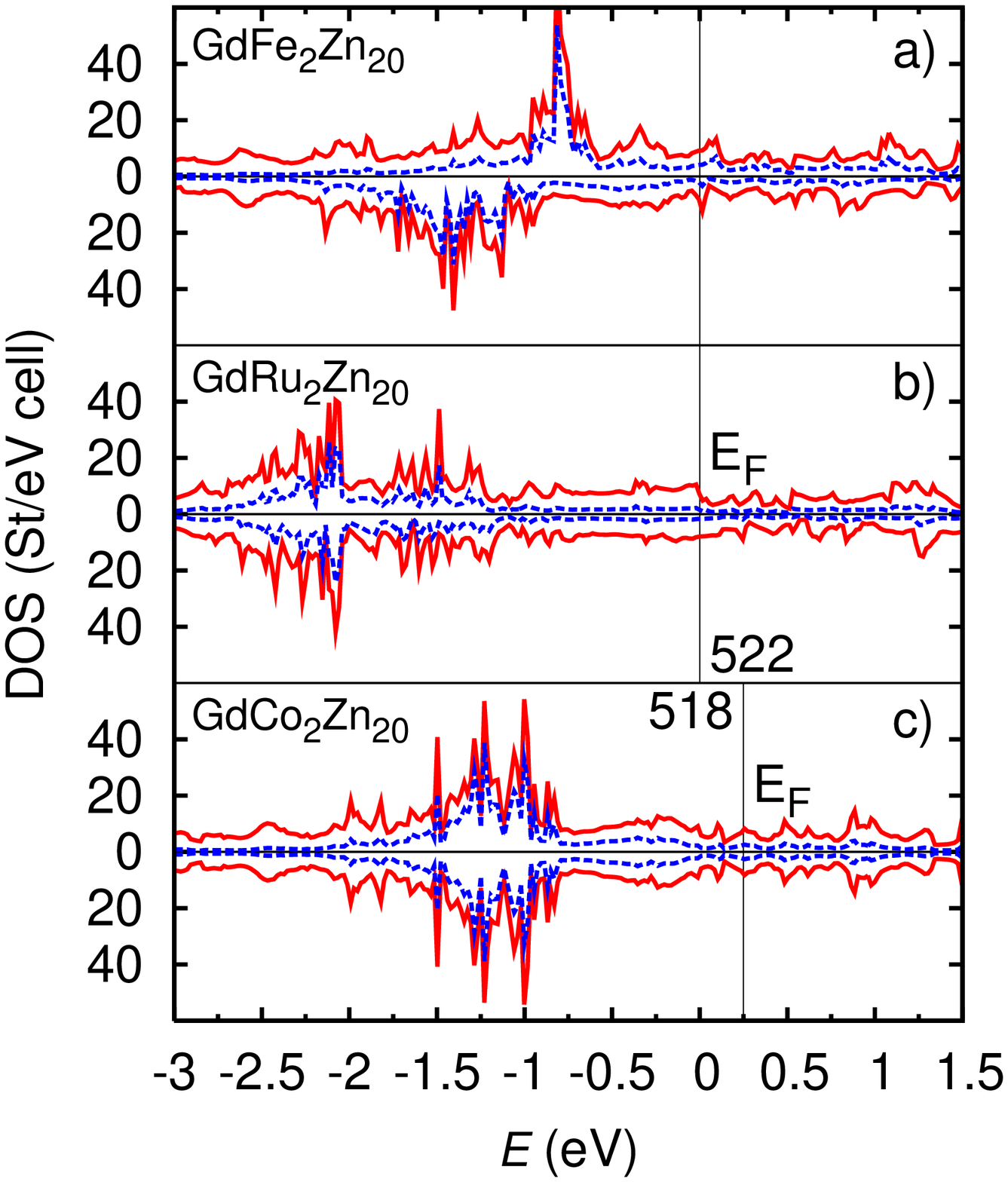}\\
  \caption{(color online) The red solid line corresponds to DOS of FM-ordered GdFe$_2$Zn$_{20}$ 
(a), FM-ordered GdRu$_2$Zn$_{20}$ (b) and AFM one GdCo$_2$Zn$_{20}$ (c) near
$E_F$ (in St/eV cell) and partial DOS of Fe, Ru and Co atoms (blue dashed line)
(in St/eV atom). $E_F$ is shown by vertical lines. 518 and 522 corresponds to number of valence
electrons in the unit cell calculated in the rigid band approximation from the DOS.}
  \label{Fig22DOSGdall}
  \end{center}
\end{figure}

The electronic structure calculation of the three GdT$_2$Zn$_{20}$ analogues, based on the treatment of $4f$ electrons in core states, can help to understand the effect of a submerging Gd$^{3+}$ local moment in these electronic backgrounds (Y analogues).
Our calculations demonstrate that, in the ordered state, Gd and the transition metal carry magnetic moments (see Table \ref{table4}).
Magnetic moments of Gd atoms are about $7.4~\mu _B$ for FM ordered compounds and $7.3~\mu _B$ for AFM ordered compound, significantly smaller compared to elemental Gd result\cite{perlov_rare, turek_ab}, $7.6~\mu _B$.
The magnetic moment additional to the Hund's value ($7~\mu _B$) comes from the polarization of Gd's $p$ and $d$ states by magnetic $4f$ electrons.
The negative coupling between Gd and transition metals induces magnetic moments on these atoms in direction opposite to the Gd magnetic moment.
In agreement with the high DOS of Fe atoms in YFe$_2$Zn$_{20}$, the induced magnetic moment on Fe atoms, $-0.84~\mu _B$, is the largest among all series.
The smaller DOS of Ru atoms in YRu$_2$Zn$_{20}$ compound correlates with a smaller induced magnetic moment on Ru in GdRu$_2$Zn$_{20}$: $-0.04~\mu _B$.
The induced magnetic moment on Co is zero because of the compensation of interactions with Gd in AFM GdCo$_2$Zn$_{20}$. 
The calculated total magnetic moment, $7.25~\mu _B$, $6.53~\mu _B$ and $7.30~\mu _B$ for GdT$_2$Zn$_{20}$ (T = Co, Fe and Ru respectively), are in good agreement with the experimental values, $7.3~\mu _B$, $6.5~\mu _B$ and $7.25~\mu _B$ (see Table \ref{table2}). 
The DOS for GdFe$_2$Zn$_{20}$ [Fig. \ref{Fig22DOSGdall}(a)] demonstrates a significant splitting between occupied and empty $3d$ states of Fe, in agreement with sizable Fe magnetic moments, whereas this splitting is almost negligible in case of Ru based compounds [Fig. \ref{Fig22DOSGdall}(b)].

\begin{table}
\caption{
The calculated DOS in St/eV cell ($N(E_F)$), averaged DOS per one atom 
($N(E_F)/N_{atoms}$), partial DOS at transition metal atom ($N_{T}(E_F)$) and magnetic moment in $\mu_B$ for Gd and transition metal, T, in GdT$_2$Zn$_{20}$ compounds.
}
\label{table4}
\begin{ruledtabular}
\begin{tabular}{lccccc}
Compound & $N(E_F)$ & $N(E_F)/N_{atoms}$ & $N_{T}(E_F)$&\multicolumn{2}{c}{Magnetic Moment} \\
\cline{5-6}
      &   & & & Gd & T \\
\hline
Pt (elemental)              & 2.2 & 2.2 & 2.2 & &  \\
\hline
Pd (elemental)             & 2.6 & 2.6 & 2.6 & &  \\
\hline
Fe (elemental)             & 3.5 & 3.5 & 3.5 & &  \\
\hline
YCo$_2$Zn$_{20}$ & 16.32 &  0.35 & 1.28 & &   \\
\hline
YFe$_2$Zn$_{20}$ & 31.35 &  0.68 & 2.86 &  &   \\
\hline
YRu$_2$Zn$_{20}$ & 18.72 &  0.41 & 1.0 &  &   \\
\hline
GdCo$_2$Zn$_{20}$ & 14.92 &  & & 7.25 &  0.00 \\
\hline
GdFe$_2$Zn$_{20}$ & 17.95 &  & & 7.37 & -0.84 \\
\hline
GdRu$_2$Zn$_{20}$ & 17.15 &  & & 7.34 & -0.04 \\
\end{tabular}
\end{ruledtabular}
\end{table}

\subsection{\label{sec:E}Gd(Fe$_x$Co$_{1-x}$)$_2$Zn$_{20}$ and Y(Fe$_x$Co$_{1-x}$)$_2$Zn$_{20}$}

\begin{figure}
  \begin{center}
  \includegraphics[clip, width=0.45\textwidth]{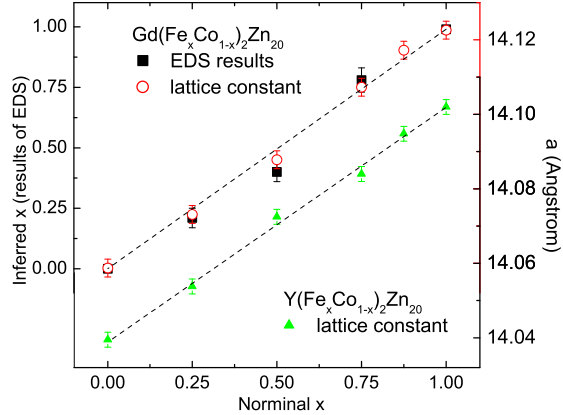}\\
  \caption{(color online) Lattice constants of the series of Gd(Fe$_x$Co$_{1-x}$)$_2$Zn$_{20}$ (open circle) and Y(Fe$_x$Co$_{1-x}$)$_2$Zn$_{20}$ (solid triangle). Fe concentration of Gd(Fe$_x$Co$_{1-x}$)$_2$Zn$_{20}$ series inferred from EDS measurements (solid square).}
  \label{Fig23EDS}
  \end{center}
\end{figure}

Based on the distinct difference between the RFe$_2$Zn$_{20}$ and RCo$_2$Zn$_{20}$ compounds and motivated by the band structural calculations, a systematic study of R(Fe$_x$Co$_{1-x}$)$_2$Zn$_{20}$ for R = Gd and Y was made.
The same growth conditions for T = Fe and Co samples further facilitates such a study of the effects of $3d$ band filling, as well as proximity to the Stoner limit, on the magnetic ordering found in GdFe$_2$Zn$_{20}$.
In order to check $x$ of Gd(Fe$_x$Co$_{1-x}$)$_2$Zn$_{20}$ and Y(Fe$_x$Co$_{1-x}$)$_2$Zn$_{20}$, Energy Dispersive Spectra (EDS) measurements, a direct method to determine the elements concentrations, and powder X-ray diffraction measurements were employed.
Figure \ref{Fig23EDS} presents EDS measurement results for the Gd series, and the lattice constants for both series.
The linear variation of lattice constants with $x$ for both series is compliant with Vegard's law, which is consistent with the results of EDS.
Due to these results; the nominal $x$ value is used from this point onward.

\begin{figure}
  \begin{center}
  \includegraphics[clip, width=0.45\textwidth]{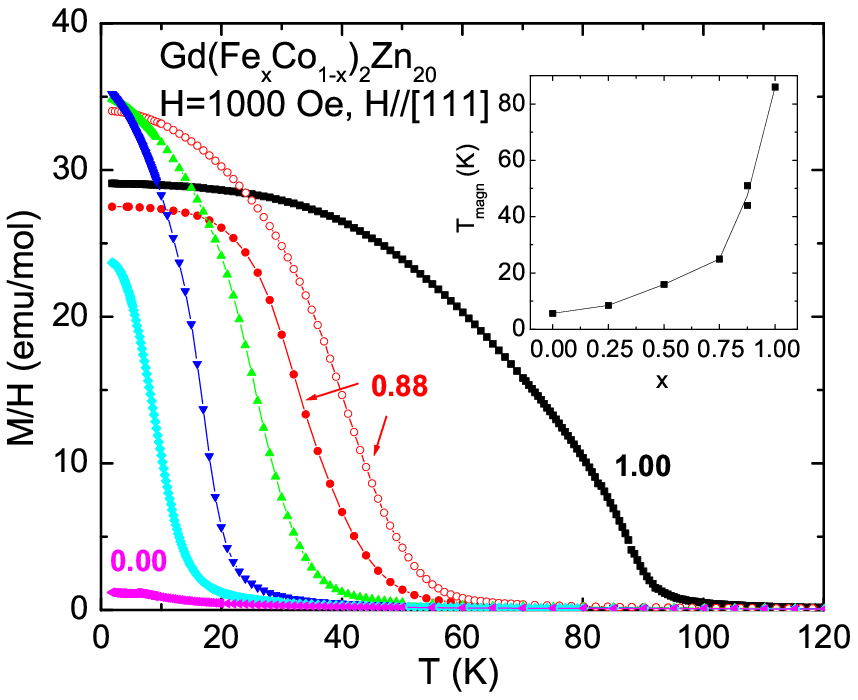}\\
  \caption{(color online) $M/H$ of Gd(Fe$_x$Co$_{1-x}$)$_2$Zn$_{20}$ series versus temperature for $x = 1.00$, $0.88$, $0.75$, $0.50$, $0.25$ and $0$ from right to left. Note data from two samples of $x = 0.88$ are shown. Inset: magnetic phase transition temperatures for Gd(Fe$_x$Co$_{1-x}$)$_2$Zn$_{20}$. }
  \label{Fig24GdFeCoMT}
  \end{center}
\end{figure}

\begin{figure}
  \begin{center}
  \includegraphics[clip, width=0.45\textwidth]{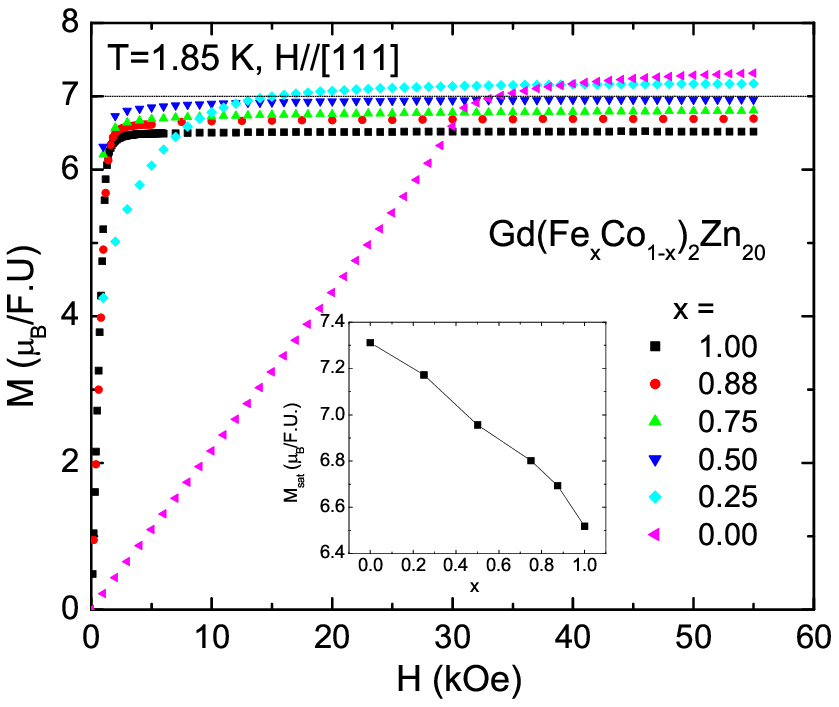}\\
  \caption{(color online) Low temperature ($T = 1.85$~K) magnetization versus applied field for the series of Gd(Fe$_x$Co$_{1-x}$)$_2$Zn$_{20}$. Inset: saturated moments as a function of $x$.}
  \label{Fig25GdFeCoMH}
  \end{center}
\end{figure}

Figure \ref{Fig24GdFeCoMT} shows the magnetization divided by the applied field as a function of temperature for Gd(Fe$_x$Co$_{1-x}$)$_2$Zn$_{20}$, which indicates FM ordering for $x \geq  0.25$.
As $x$ is increased from 0 to 1, the polarizability of the electronic background [Y(Fe$_x$Co$_{1-x}$)$_2$Zn$_{20}$] increases, and there is a monotonic, but super-linear increase in $T_{mag}$ (inset of Fig. \ref{Fig24GdFeCoMT}), which is, reminiscent of the $x$ dependent of $Z$ inferred from measurements on Y(Fe$_x$Co$_{1-x}$)$_2$Zn$_{20}$.\cite{jia_nearly_2007}
Clearer evidence of the FM ground states for $x\geq 0.25$ is present in the low temperature magnetization isotherms (Fig. \ref{Fig25GdFeCoMH}).
The saturated moment extracted from the magnetization values, under 55 kOe applied field along [111] crystallographic direction, varies monotonically from the slightly enhanced value of $7.3~\mu _B$ for GdCo$_2$Zn$_{20}$ to the slightly deficient value of $6.5~\mu _B$ for GdFe$_2$Zn$_{20}$.

\section{Discussion}

The band structure calculation indicates that, with same structure and similar lattice parameters, the diverse magnetic properties of GdT$_2$Zn$_{20}$ and YT$_2$Zn$_{20}$ are mainly dependent on the $d$-band conduction electrons from the transition metal site.
The different $d$-band filling of the Fe column members and the Co column members is associated with the different sign of the magnetic coupling of Gd$^{3+}$ local moments, and thereupon the different type of the magnetic ordering.
Furthermore, the high and intermediately high $N(E_F)$ of $3d$ and $4d$ sub-bands of Fe and Ru, respectively, are associated with the strongly correlated electronic state of YFe$_2$Zn$_{20}$ and YRu$_2$Zn$_{20}$, as well as the strong coupling between the Gd$^{3+}$ local moments in GdFe$_2$Zn$_{20}$ and GdRu$_2$Zn$_{20}$, and therefore the high magnetic ordering temperatures.
The negative induced moment on Fe site is not unexpected in intermetallic systems consisting of a heavy rare earth and an over-half-filled, $3d$ transition metal\cite{franse_magnetic_1993, brooks_density_1993}, which can be understood in terms of the hybridization between the $3d$ electrons of transition metal and the $5d$ electrons of the rare earth\cite{campbell_indirect_1972}. 

In addition to the electronic structure calculation, the remarkable high-temperature FM ordering of GdFe$_2$Zn$_{20}$ and GdRu$_2$Zn$_{20}$ can be understood in the conceptually simple context of the large Heisenberg moments, associated with the Gd$^{3+}$ ion ($\textbf {S} = 7/2$), being embedded in the NFFL associated with YFe$_2$Zn$_{20}$ and YRu$_2$Zn$_{20}$.
This framework has been employed to understand the anomalously high temperature FM ordering in some systems of local moments in NFFL hosts, such as dilute Fe, Co, or Gd in Pd or Pt\cite{nieuwenhuys_magnetic_1975, crangle_ferromagnetism_1964}, or RCo$_2$(R = Gd - Tm)\cite{duc_itinerant_1999, duc_formation_1999}.  
In these systems, the itinerant electrons of the host (Pd, Pt or YCo$_2$) are polarized by the local moments (Fe, Co ions or R$^{3+}$ ions), strongly couple them, and result in high-temperature, local moments ordering and induced moment of themselves.

The substitutional series of Gd(Fe$_x$Co$_{1-x}$)$_2$Zn$_{20}$ and Y(Fe$_x$Co$_{1-x}$)$_2$Zn$_{20}$ provide the versatility to study the correlation between the local moments and the high polarizable host.
When $x$ is varied from 0 to 1, by tuning of the $d$-band filling, the inferred values of $Z$ for the Y(Fe$_x$Co$_{1-x}$)$_2$Zn$_{20}$ series, representing to some extent the polarizability, increase super-lineally from 0.43 to 0.88,\cite{jia_nearly_2007} giving rise to the highly non-linear increase of the magnetic ordering temperature for the Gd(Fe$_x$Co$_{1-x}$)$_2$Zn$_{20}$ series (Fig. \ref{Fig24GdFeCoMT}).
This correspondence between the $Z$ values and the magnetic ordering temperatures is even consistent with the $T_{\mathrm{C}}$ value for GdRu$_2$Zn$_{20}$, although the itinerant electrons of the transition metal are $4d$, not $3d$.
Given $Z=0.67$ for YRu$_2$Zn$_{20}$, a similar $Z$-value of the host is between $x=0.5$ and $0.75$ for Y(Fe$_x$Co$_{1-x}$)$_2$Zn$_{20}$.\cite{jia_nearly_2007} The $T_{\mathrm{C}}$ value of GdRu$_2$Zn$_{20}$ is also between the $T_{\mathrm{C}}$ values for $x=0.5$ and $0.75$ Gd(Fe$_x$Co$_{1-x}$)$_2$Zn$_{20}$.   
   
\begin{figure}
  \begin{center}
  \includegraphics[clip, width=0.45\textwidth]{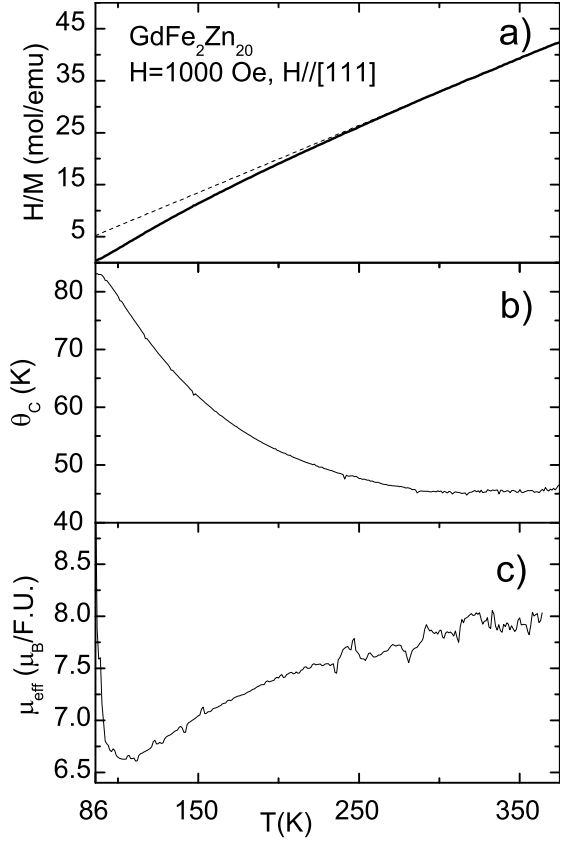}\\
  \caption{(a): $H/M$ ($H = 1000$~kOe) of GdFe$_2$Zn$_{20}$ as a function of temperature. The dash line represents the Curie-Weiss fit above 250 K. (b): temperature varied $\theta _C$. (c): temperature varied $\mu _{eff}$. (See text)}
  \label{Fig26GdFeHM}
  \end{center}
\end{figure}

This conceptually simple framework can also help to understand the curious temperature dependence of the $1/\chi(T)$ data for GdFe$_2$Zn$_{20}$.
Figure \ref{Fig26GdFeHM}(a) shows the temperature dependent $H/M$ in applied field ($H = 1000$ Oe), with a dash line presenting the CW fit above 250 K.
As shown before, the fit gives the value of the effective moment ($\mu _{eff}=7.9~\mu _B$), comparable with the effective moment of $4f$ electrons of Gd$^{3+}$ in Hund's ground state.
The deviation from the CW law below 250 K has been explained as a result of temperature dependent coupling between Gd$^{3+}$ local moments by means of strongly polarizable electronic background\cite{jia_nearly_2007}.
Assuming a constant $\mu _{eff}$, one can extract the temperature dependence of $\theta _C$ from the $1/\chi$ data.
As shown in Fig. \ref{Fig26GdFeHM}(b), $\theta _C$ essentially constant ($\sim 45$ K) above 275 K; then increases monotonically as temperature decrease, tracking $\chi (T)$ of YFe$_2$Zn$_{20}$ (Fig. \ref{Fig17YMT}).

\begin{figure}
  \begin{center}
  \includegraphics[clip, width=0.45\textwidth]{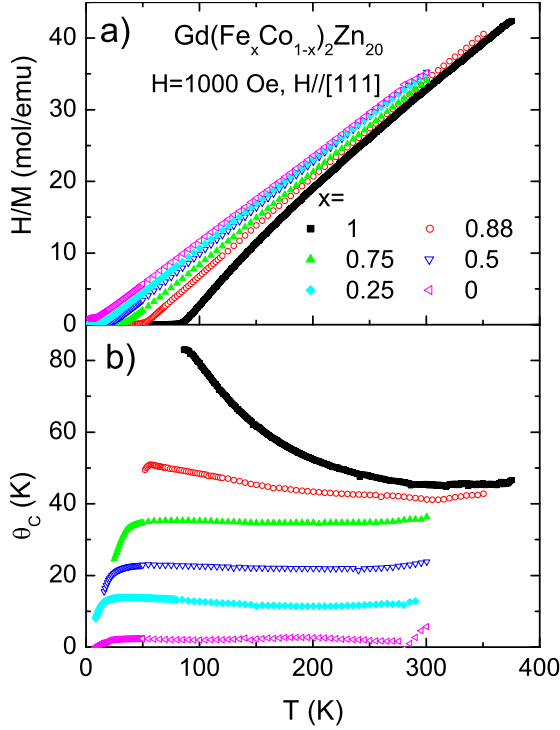}\\
  \caption{(color online) (a): $H/M$ ($H = 1000$~kOe) of Gd(Fe$_x$Co$_{1-x}$)$_2$Zn$_{20}$ as a function of temperature. (b): temperature varied $\theta _C$. (See text)}
  \label{Fig27GdFeCoHM}
  \end{center}
\end{figure}

The correlation of the temperature dependent $\chi$ and the polarizability of electronic background, can also be seen in the susceptibility of Gd(Fe$_x$Co$_{1-x}$)$_2$Zn$_{20}$ series.
Figure \ref{Fig27GdFeCoHM}(a) presents temperature dependent $H/M$ under the applied field $H = 1000$ Oe.
Linear and parallel to each other at high temperature region, the data sets start to deviate at lower temperature, especially for large $x$.
Similar to discussed before, the temperature dependent $\theta _C$ values were extracted with the assumption of invariant $\mu _{eff}$.
Figure \ref{Fig27GdFeCoHM}(b) shows that $\theta _C$ varies strongly, much weakly and negligibly as $x = 1$, $0.88$ and $\leq 0.75$, respectively.
For each $x$, the variation of $\theta _C$ tracks $\chi(T)$ of the Y(Fe$_x$Co$_{1-x}$)$_2$Zn$_{20}$ series.\cite{jia_nearly_2007}

An alternative method of analyzing the $\chi (T)$ data assumes that some induced moment exists above $T_{\mathrm{C}}$ and is aligned locally anti-parallel to the Gd moment (in essence forming a composite moment).
Assuming an invariant $\theta _C$, values of $C$ can be inferred from:

\begin{equation}
1/C\approx \frac{d(\frac{T-\theta _C}{C})}{dT}=\frac{d(\frac{H}{M})}{dT}.
\label{eqn:3}
\end{equation}

Shown in Fig. \ref{Fig26GdFeHM}(c), $\mu _{eff}$ manifests a monotonic decrease with decreasing temperature down to 110 K, at which temperature it shows a minimum value $6.6~\mu _B$.
From 100 K to $T_{\mathrm{C}}$, $\mu _{eff}$ starts to rise with a highly non-linear fashion.
This rise of the $\mu _{eff}$ value is not unexpected in the vicinity of $T_{\mathrm{C}}$ in FM system\cite{mydosh_spin_1993}, and could be due to the short range ordering or formation of magnetic clusters of the Gd$^{3+}$ local moment and induced moment.
The decrease of $\mu _{eff}$, in this scenario, would be the result of the formation of the magnetic droplets, consisted with the Gd$^{+3}$ local moments and the oppositely polarized electron cloud from the highly polarizable host.
Such magnetic droplets are not unprecidented in analogous systems, above $T_{\mathrm{C}}$.
For example, the `giant moment' was observed in dilute Fe-Pd alloy\cite{clogston_local_1962}; the deficient $\mu _{eff}$ of local moments was also found in RCo$_2$ series (R = Gd-Tm)\cite{stewart_paramagnetic_1984} above $T_{\mathrm{C}}$.
Giving that the primary difference between these two alternative explanations is whether the itinerant electrons are polarized above $T_{\mathrm{C}}$, M\"{o}ssbauer spectra measurements on the Fe sites at varied temperature can resolve this paradox.

\section{Summary}

The six GdT$_2$Zn$_{20}$ (T = Fe, Ru, Os, Co, Rh and Ir) compounds have magnetic properties that differ dramatically between the Fe column and the Co column members.
The Fe column members order ferromagnetically with the enhanced transition temperatures for the T = Fe and Ru members, whereas the Co column members all manifest low-temperature, AFM ordering.
In a related manner, the T = Fe and Ru members of YT$_2$Zn$_{20}$ family manifest typical properties associated with NFFLs.
Band structure calculation results for the T = Fe and Ru members reveal that the large DOS at the Fermi level is correlated with the enhancement in the their magnetic properties.
The data on the pseudo-ternary series of compounds Gd(Fe$_x$Co$_{1-x}$)$_2$Zn$_{20}$ and Y(Fe$_x$Co$_{1-x}$)$_2$Zn$_{20}$ further display the effect of the different $3d$-band filling on the magnetic properties of these two series.
The conceptually simple framework of the Heisenberg moments embedded in the NFFL, was discussed to understand the enhanced transitions for GdFe$_2$Zn$_{20}$ and GdRu$_2$Zn$_{20}$ and the curious temperature dependence of the $1/\chi$ versus $T$ data for GdFe$_2$Zn$_{20}$.

\begin{acknowledgments}
The authors thank J. Frederich for growing some of the compounds, L. Tan for Laue X-ray measurements, E. D. Mun, X. Zong and R. Prozorov for helpful discussions.
SLB thanks David Lodge for valuable insights.
Ames Laboratory is operated for the U.S. Department of Energy by Iowa State University under Contract No. DE-AC02-07CH11358.
This work was supported by the Director for Energy Research, Office of Basic Energy Sciences. 

\end{acknowledgments}



\begin{thebibliography}{44}
\expandafter\ifx\csname natexlab\endcsname\relax\def\natexlab#1{#1}\fi
\expandafter\ifx\csname bibnamefont\endcsname\relax
  \def\bibnamefont#1{#1}\fi
\expandafter\ifx\csname bibfnamefont\endcsname\relax
  \def\bibfnamefont#1{#1}\fi
\expandafter\ifx\csname citenamefont\endcsname\relax
  \def\citenamefont#1{#1}\fi
\expandafter\ifx\csname url\endcsname\relax
  \def\url#1{\texttt{#1}}\fi
\expandafter\ifx\csname urlprefix\endcsname\relax\def\urlprefix{URL }\fi
\providecommand{\bibinfo}[2]{#2}
\providecommand{\eprint}[2][]{\url{#2}}

\bibitem[{\citenamefont{Franse and Radwanski}(1993)}]{franse_magnetic_1993}
\bibinfo{author}{\bibfnamefont{J.~J.~M.} \bibnamefont{Franse}}
  \bibnamefont{and} \bibinfo{author}{\bibfnamefont{R.~J.}
  \bibnamefont{Radwanski}}, \emph{\bibinfo{title}{\rm{in} \it{Handbook of
  Magnetic Materials} \rm{vol. 7 Edited by K.H.J. Buschow}}}
  (\bibinfo{publisher}{Amsterdam: Elsevier}, \bibinfo{year}{1993}), pp.
  \bibinfo{pages}{307--501}.

\bibitem[{\citenamefont{Szytula and Leciejewicz}(1994)}]{szytula_handbook_1994}
\bibinfo{author}{\bibfnamefont{A.}~\bibnamefont{Szytula}} \bibnamefont{and}
  \bibinfo{author}{\bibfnamefont{J.}~\bibnamefont{Leciejewicz}},
  \emph{\bibinfo{title}{Handbook of Crystal Structures and Magnetic Properties
  of Rare Earth Intermetallics}} (\bibinfo{publisher}{CRC Press},
  \bibinfo{year}{1994}).

\bibitem[{\citenamefont{Jia et~al.}(2007{\natexlab{a}})\citenamefont{Jia,
  Bud'ko, Samolyuk, and Canfield}}]{jia_nearly_2007}
\bibinfo{author}{\bibfnamefont{S.}~\bibnamefont{Jia}},
  \bibinfo{author}{\bibfnamefont{S.~L.} \bibnamefont{Bud'ko}},
  \bibinfo{author}{\bibfnamefont{G.~D.} \bibnamefont{Samolyuk}},
  \bibnamefont{and} \bibinfo{author}{\bibfnamefont{P.~C.}
  \bibnamefont{Canfield}}, \bibinfo{journal}{Nat Phys}
  \textbf{\bibinfo{volume}{3}}, \bibinfo{pages}{334}
  (\bibinfo{year}{2007}{\natexlab{a}}).

\bibitem[{\citenamefont{Torikachvili et~al.}(2007)\citenamefont{Torikachvili,
  Jia, Mun, Hannahs, Black, Neils, Martien, Bud'ko, and
  Canfield}}]{torikachvili_six_2007}
\bibinfo{author}{\bibfnamefont{M.~S.} \bibnamefont{Torikachvili}},
  \bibinfo{author}{\bibfnamefont{S.}~\bibnamefont{Jia}},
  \bibinfo{author}{\bibfnamefont{E.~D.} \bibnamefont{Mun}},
  \bibinfo{author}{\bibfnamefont{S.~T.} \bibnamefont{Hannahs}},
  \bibinfo{author}{\bibfnamefont{R.~C.} \bibnamefont{Black}},
  \bibinfo{author}{\bibfnamefont{W.~K.} \bibnamefont{Neils}},
  \bibinfo{author}{\bibfnamefont{D.}~\bibnamefont{Martien}},
  \bibinfo{author}{\bibfnamefont{S.~L.} \bibnamefont{Bud'ko}},
  \bibnamefont{and} \bibinfo{author}{\bibfnamefont{P.~C.}
  \bibnamefont{Canfield}}, \bibinfo{journal}{PNAS}
  \textbf{\bibinfo{volume}{104}}, \bibinfo{pages}{9960} (\bibinfo{year}{2007}).

\bibitem[{\citenamefont{Jia et~al.}(2007{\natexlab{b}})\citenamefont{Jia, Ni,
  Bud'ko, and Canfield}}]{jia_GdY_2007}
\bibinfo{author}{\bibfnamefont{S.}~\bibnamefont{Jia}},
  \bibinfo{author}{\bibfnamefont{N.}~\bibnamefont{Ni}},
  \bibinfo{author}{\bibfnamefont{S.~L.} \bibnamefont{Bud'ko}},
  \bibnamefont{and} \bibinfo{author}{\bibfnamefont{P.~C.}
  \bibnamefont{Canfield}}, \bibinfo{journal}{Physical Review B (Condensed
  Matter and Materials Physics)} \textbf{\bibinfo{volume}{76}},
  \bibinfo{eid}{184410} (\bibinfo{year}{2007}{\natexlab{b}}).

\bibitem[{\citenamefont{Nasch et~al.}(1997)\citenamefont{Nasch, Jeitschko, and
  Rodewald}}]{nasch_ternary_1997}
\bibinfo{author}{\bibfnamefont{T.}~\bibnamefont{Nasch}},
  \bibinfo{author}{\bibfnamefont{W.}~\bibnamefont{Jeitschko}},
  \bibnamefont{and} \bibinfo{author}{\bibfnamefont{U.~C.}
  \bibnamefont{Rodewald}}, \bibinfo{journal}{Zeitschrift fuer Naturforschung,
  B: Chemical Sciences} \textbf{\bibinfo{volume}{52}}, \bibinfo{pages}{1023}
  (\bibinfo{year}{1997}).

\bibitem[{\citenamefont{Kripyakevich and
  Zarechnyuk}(1968)}]{kripyakevich_RCr2Al20_1968}
\bibinfo{author}{\bibfnamefont{P.~I.} \bibnamefont{Kripyakevich}}
  \bibnamefont{and} \bibinfo{author}{\bibfnamefont{O.~S.}
  \bibnamefont{Zarechnyuk}}, \bibinfo{journal}{Dopov. Akad. Nauk Ukr. RSR, Ser.
  A} \textbf{\bibinfo{volume}{30}}, \bibinfo{pages}{364}
  (\bibinfo{year}{1968}).

\bibitem[{\citenamefont{Thiede et~al.}(1998)\citenamefont{Thiede, Jeitschko,
  Niemann, and Ebel}}]{thiede_euta2al20_1998}
\bibinfo{author}{\bibfnamefont{V.~M.~T.} \bibnamefont{Thiede}},
  \bibinfo{author}{\bibfnamefont{W.}~\bibnamefont{Jeitschko}},
  \bibinfo{author}{\bibfnamefont{S.}~\bibnamefont{Niemann}}, \bibnamefont{and}
  \bibinfo{author}{\bibfnamefont{T.}~\bibnamefont{Ebel}},
  \bibinfo{journal}{Journal of Alloys and Compounds}
  \textbf{\bibinfo{volume}{267}}, \bibinfo{pages}{23} (\bibinfo{year}{1998}).

\bibitem[{\citenamefont{Moze et~al.}(1998)\citenamefont{Moze, Tung, Franse, and
  Buschow}}]{moze_crystal_1998}
\bibinfo{author}{\bibfnamefont{O.}~\bibnamefont{Moze}},
  \bibinfo{author}{\bibfnamefont{L.~D.} \bibnamefont{Tung}},
  \bibinfo{author}{\bibfnamefont{J.~J.~M.} \bibnamefont{Franse}},
  \bibnamefont{and} \bibinfo{author}{\bibfnamefont{K.~H.~J.}
  \bibnamefont{Buschow}}, \bibinfo{journal}{Journal of Alloys and Compounds}
  \textbf{\bibinfo{volume}{268}}, \bibinfo{pages}{39} (\bibinfo{year}{1998}).

\bibitem[{\citenamefont{Canfield and Fisk}(1992)}]{canfield_growth_1992}
\bibinfo{author}{\bibfnamefont{P.~C.} \bibnamefont{Canfield}} \bibnamefont{and}
  \bibinfo{author}{\bibfnamefont{Z.}~\bibnamefont{Fisk}},
  \bibinfo{journal}{Philosophical Magazine Part B}
  \textbf{\bibinfo{volume}{65}}, \bibinfo{pages}{1117} (\bibinfo{year}{1992}).

\bibitem[{\citenamefont{Stoe}(2002)}]{stoe_area-software_2002}
\bibinfo{author}{\bibfnamefont{X.}~\bibnamefont{Stoe}},
  \emph{\bibinfo{title}{AREA-Software Suite for the STOE IPDS II}}
  (\bibinfo{publisher}{Stoe \& Cie GmbH, Darmstadt, Germany},
  \bibinfo{year}{2002}).

\bibitem[{\citenamefont{Sheldrick and SHELXTL}(2000)}]{sheldrick_2000}
\bibinfo{author}{\bibfnamefont{G.~M.} \bibnamefont{Sheldrick}}
  \bibnamefont{and} \bibinfo{author}{\bibfnamefont{D.}~\bibnamefont{SHELXTL}},
  \bibinfo{journal}{Inc., Madison, WI, USA}  (\bibinfo{year}{2000}).

\bibitem[{\citenamefont{Eiling and Schilling}(1981)}]{Pb_pressure}
\bibinfo{author}{\bibfnamefont{A.}~\bibnamefont{Eiling}} \bibnamefont{and}
  \bibinfo{author}{\bibfnamefont{J.~S.} \bibnamefont{Schilling}},
  \bibinfo{journal}{Journal of Physics F: Metal Physics}
  \textbf{\bibinfo{volume}{11}}, \bibinfo{pages}{623} (\bibinfo{year}{1981}).

\bibitem[{\citenamefont{Bud'ko et~al.}(2005)\citenamefont{Bud'ko, Wilke, Angst,
  and Canfield}}]{budko_pressure}
\bibinfo{author}{\bibfnamefont{S.}~\bibnamefont{Bud'ko}},
  \bibinfo{author}{\bibfnamefont{R.}~\bibnamefont{Wilke}},
  \bibinfo{author}{\bibfnamefont{M.}~\bibnamefont{Angst}}, \bibnamefont{and}
  \bibinfo{author}{\bibfnamefont{P.}~\bibnamefont{Canfield}},
  \bibinfo{journal}{Physica C: Superconductivity}
  \textbf{\bibinfo{volume}{420}}, \bibinfo{pages}{83} (\bibinfo{year}{2005}).

\bibitem[{\citenamefont{Chikazumi and Graham}(1997)}]{chikazumi_physics_1997}
\bibinfo{author}{\bibfnamefont{S.}~\bibnamefont{Chikazumi}} \bibnamefont{and}
  \bibinfo{author}{\bibfnamefont{C.}~\bibnamefont{Graham}},
  \emph{\bibinfo{title}{Physics of Ferromagnetism}} (\bibinfo{publisher}{Oxford
  University Press}, \bibinfo{year}{1997}).

\bibitem[{\citenamefont{Andersen}(1975)}]{andersen_linear_1975}
\bibinfo{author}{\bibfnamefont{O.~K.} \bibnamefont{Andersen}},
  \bibinfo{journal}{Phys. Rev. B} \textbf{\bibinfo{volume}{12}},
  \bibinfo{pages}{3060} (\bibinfo{year}{1975}).

\bibitem[{\citenamefont{Andersen and Jepsen}(1984)}]{andersen_explicit_1984}
\bibinfo{author}{\bibfnamefont{O.~K.} \bibnamefont{Andersen}} \bibnamefont{and}
  \bibinfo{author}{\bibfnamefont{O.}~\bibnamefont{Jepsen}},
  \bibinfo{journal}{Phys. Rev. Lett.} \textbf{\bibinfo{volume}{53}},
  \bibinfo{pages}{2571} (\bibinfo{year}{1984}).

\bibitem[{\citenamefont{von Barth and Hedin}(1972)}]{Barth_local_1972}
\bibinfo{author}{\bibfnamefont{U.}~\bibnamefont{von Barth}} \bibnamefont{and}
  \bibinfo{author}{\bibfnamefont{L.}~\bibnamefont{Hedin}},
  \bibinfo{journal}{Journal of Physics C: Solid State Physics}
  \textbf{\bibinfo{volume}{5}}, \bibinfo{pages}{1629} (\bibinfo{year}{1972}).

\bibitem[{\citenamefont{Perdew et~al.}(1996)\citenamefont{Perdew, Burke, and
  Ernzerhof}}]{perdew_generalized_1996}
\bibinfo{author}{\bibfnamefont{J.~P.} \bibnamefont{Perdew}},
  \bibinfo{author}{\bibfnamefont{K.}~\bibnamefont{Burke}}, \bibnamefont{and}
  \bibinfo{author}{\bibfnamefont{M.}~\bibnamefont{Ernzerhof}},
  \bibinfo{journal}{Phys. Rev. Lett.} \textbf{\bibinfo{volume}{77}},
  \bibinfo{pages}{3865} (\bibinfo{year}{1996}).

\bibitem[{\citenamefont{Perlov et~al.}(2000)\citenamefont{Perlov, Halilov, and
  Eschrig}}]{perlov_rare}
\bibinfo{author}{\bibfnamefont{A.~Y.} \bibnamefont{Perlov}},
  \bibinfo{author}{\bibfnamefont{S.~V.} \bibnamefont{Halilov}},
  \bibnamefont{and} \bibinfo{author}{\bibfnamefont{H.}~\bibnamefont{Eschrig}},
  \bibinfo{journal}{Phys. Rev. B} \textbf{\bibinfo{volume}{61}},
  \bibinfo{pages}{4070} (\bibinfo{year}{2000}).

\bibitem[{\citenamefont{I~Turek and Blugel}(2003)}]{turek_ab}
\bibinfo{author}{\bibfnamefont{G.~B.} \bibnamefont{I~Turek},
  \bibfnamefont{J~Kudrnovsky}} \bibnamefont{and}
  \bibinfo{author}{\bibfnamefont{S.}~\bibnamefont{Blugel}},
  \bibinfo{journal}{Journal of Physics: Condensed Matter}
  \textbf{\bibinfo{volume}{15}}, \bibinfo{pages}{2771} (\bibinfo{year}{2003}).

\bibitem[{per()}]{periodic_table}
\bibinfo{note}{From \it table of periodic properties of the elements \rm
  (Sargent-Welch/VWR Scientific Products, 1998)}.

\bibitem[{\citenamefont{Arrott}(1957)}]{arrott_criterion_1957}
\bibinfo{author}{\bibfnamefont{A.}~\bibnamefont{Arrott}},
  \bibinfo{journal}{Physical Review} \textbf{\bibinfo{volume}{108}},
  \bibinfo{pages}{1394} (\bibinfo{year}{1957}).

\bibitem[{\citenamefont{Brommer and Franse}(1990)}]{brommer_strongly_1990}
\bibinfo{author}{\bibfnamefont{P.~E.} \bibnamefont{Brommer}} \bibnamefont{and}
  \bibinfo{author}{\bibfnamefont{J.~J.~M.} \bibnamefont{Franse}},
  \emph{\bibinfo{title}{\rm{in} \it{Ferromagnetic Materials} \rm{vol 5, edited
  by K.H.J. Buschow and E.P. Wohlfarth}}} (\bibinfo{publisher}{Amsterdam:
  North-Holland}, \bibinfo{year}{1990}), pp. \bibinfo{pages}{323--396}.

\bibitem[{\citenamefont{Yeung et~al.}(1986)\citenamefont{Yeung, Roshko, and
  Williams}}]{yeung_arrott-plot_1986}
\bibinfo{author}{\bibfnamefont{I.}~\bibnamefont{Yeung}},
  \bibinfo{author}{\bibfnamefont{R.~M.} \bibnamefont{Roshko}},
  \bibnamefont{and} \bibinfo{author}{\bibfnamefont{G.}~\bibnamefont{Williams}},
  \bibinfo{journal}{Phys. Rev. B} \textbf{\bibinfo{volume}{34}},
  \bibinfo{pages}{3456} (\bibinfo{year}{1986}).

\bibitem[{\citenamefont{Fisher}(1962)}]{fisher1962}
\bibinfo{author}{\bibfnamefont{M.~E.} \bibnamefont{Fisher}},
  \bibinfo{journal}{Philos. Mag.} \textbf{\bibinfo{volume}{7}},
  \bibinfo{pages}{1731} (\bibinfo{year}{1962}).

\bibitem[{\citenamefont{Fisher and Langer}(1968)}]{fisher_resistive_1968}
\bibinfo{author}{\bibfnamefont{M.~E.} \bibnamefont{Fisher}} \bibnamefont{and}
  \bibinfo{author}{\bibfnamefont{J.~S.} \bibnamefont{Langer}},
  \bibinfo{journal}{Physical Review Letters} \textbf{\bibinfo{volume}{20}},
  \bibinfo{pages}{665} (\bibinfo{year}{1968}).

\bibitem[{\citenamefont{Ziman}(1979)}]{ziman_principles_1979}
\bibinfo{author}{\bibfnamefont{J.~M.} \bibnamefont{Ziman}},
  \emph{\bibinfo{title}{Principles of the Theory of Solids}}
  (\bibinfo{publisher}{Cambridge University Press}, \bibinfo{year}{1979}).

\bibitem[{\citenamefont{Mulay and Boudreaux}(1976)}]{mulay_theory_1976}
\bibinfo{author}{\bibfnamefont{L.~N.} \bibnamefont{Mulay}} \bibnamefont{and}
  \bibinfo{author}{\bibfnamefont{E.~A.} \bibnamefont{Boudreaux}},
  \emph{\bibinfo{title}{Theory and applications of molecular diamagnetism}}
  (\bibinfo{publisher}{Wiley}, \bibinfo{year}{1976}).

\bibitem[{pdp()}]{pdptZ}
\bibinfo{note}{$\chi _0$ and $\gamma _0$ values of Pd and Pt are extracted
  from: G. S. Knapp and R. W. Jones, Physical Review B $\bf{6}$, 1761 (1972);
  B. Zellermann, A. Paintner and J. Voitl\"{a}nder, J. Phys: Condens. Matter
  $\bf{16}$, 919 (2004)}.

\bibitem[{CuA()}]{CuAgAu}
\bibinfo{note}{$\gamma _0=0.695$ mJ/molK$^2$ [from C. Kittel, \it Introduction
  to solid state physics \rm (Wiley New York, 1986)], and $\chi _{0-dia}=13.4
  \times 10^{-6}$ emu/mol (from ref. [29])}.

\bibitem[{\citenamefont{Ashcroft and Mermin}(1976)}]{ashcroft_solid_1976}
\bibinfo{author}{\bibfnamefont{N.~W.} \bibnamefont{Ashcroft}} \bibnamefont{and}
  \bibinfo{author}{\bibfnamefont{N.~D.} \bibnamefont{Mermin}},
  \emph{\bibinfo{title}{Solid state physics}} (\bibinfo{publisher}{Saunders
  College Philadelphia, Pa}, \bibinfo{year}{1976}).

\bibitem[{\citenamefont{Shimizu et~al.}(1963)\citenamefont{Shimizu, Takahashi,
  and Katsuki}}]{shimizu_pdband_1963}
\bibinfo{author}{\bibfnamefont{M.}~\bibnamefont{Shimizu}},
  \bibinfo{author}{\bibfnamefont{T.}~\bibnamefont{Takahashi}},
  \bibnamefont{and} \bibinfo{author}{\bibfnamefont{A.}~\bibnamefont{Katsuki}},
  \bibinfo{journal}{Journal of the Physical Society of Japan}
  \textbf{\bibinfo{volume}{18}}, \bibinfo{pages}{240} (\bibinfo{year}{1963}).

\bibitem[{\citenamefont{Tanaka and Harima}(1998)}]{tanaka_mass_1998}
\bibinfo{author}{\bibfnamefont{S.}~\bibnamefont{Tanaka}} \bibnamefont{and}
  \bibinfo{author}{\bibfnamefont{H.}~\bibnamefont{Harima}},
  \bibinfo{journal}{Journal of the Physical Society of Japan}
  \textbf{\bibinfo{volume}{67}}, \bibinfo{pages}{2594} (\bibinfo{year}{1998}).

\bibitem[{\citenamefont{Hayden et~al.}(1986)\citenamefont{Hayden, Lonzarich,
  and Skriver}}]{hayden_electronic_1986}
\bibinfo{author}{\bibfnamefont{S.~M.} \bibnamefont{Hayden}},
  \bibinfo{author}{\bibfnamefont{G.~G.} \bibnamefont{Lonzarich}},
  \bibnamefont{and} \bibinfo{author}{\bibfnamefont{H.~L.}
  \bibnamefont{Skriver}}, \bibinfo{journal}{Phys. Rev. B}
  \textbf{\bibinfo{volume}{33}}, \bibinfo{pages}{4977} (\bibinfo{year}{1986}).

\bibitem[{\citenamefont{Brooks and Johansson}(1993)}]{brooks_density_1993}
\bibinfo{author}{\bibfnamefont{M.~S.~S.} \bibnamefont{Brooks}}
  \bibnamefont{and}
  \bibinfo{author}{\bibfnamefont{B.}~\bibnamefont{Johansson}},
  \emph{\bibinfo{title}{\rm{in} \it{Handbook of Magnetic Materials} \rm{vol. 7
  Edited by K.H.J. Buschow}}} (\bibinfo{publisher}{Amsterdam: Elsevier},
  \bibinfo{year}{1993}), pp. \bibinfo{pages}{139--230}.

\bibitem[{\citenamefont{Campbell}(1972)}]{campbell_indirect_1972}
\bibinfo{author}{\bibfnamefont{I.~A.} \bibnamefont{Campbell}},
  \bibinfo{journal}{Journal of Physics F: Metal Physics}
  \textbf{\bibinfo{volume}{2}}, \bibinfo{pages}{L47} (\bibinfo{year}{1972}).

\bibitem[{\citenamefont{Nieuwenhuys}(1975)}]{nieuwenhuys_magnetic_1975}
\bibinfo{author}{\bibfnamefont{G.~J.} \bibnamefont{Nieuwenhuys}},
  \bibinfo{journal}{Advances in Physics} \textbf{\bibinfo{volume}{24}},
  \bibinfo{pages}{515} (\bibinfo{year}{1975}).

\bibitem[{\citenamefont{Crangle}(1964)}]{crangle_ferromagnetism_1964}
\bibinfo{author}{\bibfnamefont{J.}~\bibnamefont{Crangle}},
  \bibinfo{journal}{Physical Review Letters} \textbf{\bibinfo{volume}{13}},
  \bibinfo{pages}{569} (\bibinfo{year}{1964}).

\bibitem[{\citenamefont{Duc and Goto}(1999)}]{duc_itinerant_1999}
\bibinfo{author}{\bibfnamefont{N.~H.} \bibnamefont{Duc}} \bibnamefont{and}
  \bibinfo{author}{\bibfnamefont{T.}~\bibnamefont{Goto}},
  \emph{\bibinfo{title}{\rm{in} \it{Handbook on the Physics and Chemistry of
  Rare Earths} \rm{vol. 26, edited by K.A. Gschneidner, Jr. and L. Eyring}}}
  (\bibinfo{publisher}{Amsterdam: Elsevier}, \bibinfo{year}{1999}), pp.
  \bibinfo{pages}{177--264}.

\bibitem[{\citenamefont{Duc and Brommer}(1999)}]{duc_formation_1999}
\bibinfo{author}{\bibfnamefont{N.~H.} \bibnamefont{Duc}} \bibnamefont{and}
  \bibinfo{author}{\bibfnamefont{P.~E.} \bibnamefont{Brommer}},
  \emph{\bibinfo{title}{\rm{in} \it{Handbook of Magnetic Materials}, \rm{vol.
  12, Edited by K.H.J. Buschow}}} (\bibinfo{publisher}{Amsterdam: Elsevier},
  \bibinfo{year}{1999}), pp. \bibinfo{pages}{259--394}.

\bibitem[{\citenamefont{Mydosh}(1993)}]{mydosh_spin_1993}
\bibinfo{author}{\bibfnamefont{J.~A.} \bibnamefont{Mydosh}},
  \emph{\bibinfo{title}{Spin Glass: An Experimental Introduction}}
  (\bibinfo{publisher}{Taylor and Francis, London}, \bibinfo{year}{1993}).

\bibitem[{\citenamefont{Clogston et~al.}(1962)\citenamefont{Clogston, Matthias,
  Peter, Williams, Corenzwit, and Sherwood}}]{clogston_local_1962}
\bibinfo{author}{\bibfnamefont{A.~M.} \bibnamefont{Clogston}},
  \bibinfo{author}{\bibfnamefont{B.~T.} \bibnamefont{Matthias}},
  \bibinfo{author}{\bibfnamefont{M.}~\bibnamefont{Peter}},
  \bibinfo{author}{\bibfnamefont{H.~J.} \bibnamefont{Williams}},
  \bibinfo{author}{\bibfnamefont{E.}~\bibnamefont{Corenzwit}},
  \bibnamefont{and} \bibinfo{author}{\bibfnamefont{R.~C.}
  \bibnamefont{Sherwood}}, \bibinfo{journal}{Physical Review}
  \textbf{\bibinfo{volume}{125}}, \bibinfo{pages}{541} (\bibinfo{year}{1962}).

\bibitem[{\citenamefont{Stewart}(1984)}]{stewart_paramagnetic_1984}
\bibinfo{author}{\bibfnamefont{A.~M.} \bibnamefont{Stewart}},
  \bibinfo{journal}{Journal of Physics C: Solid State Physics}
  \textbf{\bibinfo{volume}{17}}, \bibinfo{pages}{1557} (\bibinfo{year}{1984}).

\end{thebibliography}
\end{document}